\documentclass{article}%
\usepackage{amsmath}
\usepackage{amsfonts}
\usepackage{amssymb}
\usepackage{lscape}
\usepackage{graphicx}%
\newcommand{\bpartial}{\mathop{\partial\kern -4pt\raisebox{.8pt}{$|$}}}
\newcommand{\bra}{\mathopen{[\kern-1.6pt[}}
\newcommand{\ket}{\mathclose{]\kern-1.5pt]}}
\newcommand{\bbra}{\mathopen{[\kern-2.2pt[\kern-2.3pt[}}
\newcommand{\bket}{\mathclose{]\kern-2.1pt]\kern-2.3pt]}}

\makeindex
\begin{document}
%%%%%%%%%%%%%%%%%%%%%%%%%%%%%%%%%%%%%%%%%%%%%%%%%%%%%%%%%%%%%%%%%%%%%%%%%%%%%%%%%%%%
\title{\bf Automorphism group and ad-invariant metric on all six dimensional solvable real Lie algebras }
\author { A. Rezaei-Aghdam \hspace{-3mm}{ \footnote{e-mail: rezaei-a@azaruniv.edu}\hspace{2mm}
, M. Sephid \hspace{-3mm}{ \footnote{ e-mail: s.sephid@azaruniv.edu}}\hspace{2mm}} {\small and } S. Fallahpour \\
{\small{\em Department of Physics, Faculty of science, Azarbaijan University }}\\
{\small{\em of Tarbiat Moallem , 53714-161, Tabriz, Iran  }}}
%%%%%%%%%%%%%%%%%%%%%%%%%%%%%%%%%%%%%%%%%%%%%%%%%%%%%%%%%%%%%%%%%%%%%%%%%%%%%%%%%%%%%%%%
\maketitle
%%%%%%%%%%%%%%%%%%%%%%%%%%%%%%%%%%%%%%%%%%%%%%%%%%%%%%%%%%%%%%%%%%%%%%%%%%%%%%%%%%%%%%%%
\begin{abstract}
Using adjoint representation of Lie algebras, we calculate the
automorphism group and ad-invariant metric on six dimensional
solvable real Lie algebras with 5, 4 and 3 dimensional
nilradicals.
\end{abstract}
%%%%%%%%%%%%%%%%%%%%%%%%%%%%%%%%%%%%%%%%%%%%%%%%%%%%%%%%%%%%%%%%%%%%%%%%%%%%%%%%%%%%%%%%
\newpage
\section{\bf Introduction}

The automorphisms of real three dimensional Lie algebras
\cite{Harvey} is a powerful tool for analyzing the dynamics of
3+1 dimensional Bianchi cosmological models \cite{Henneaux}.
Furthermore at the classical level, time dependent automorphism
inducing diffeomorphisms can be used to simplify the line element
and thus Einstein's field equations and also provide an algorithm
for counting the number of essential constants
(see\cite{ChrisJMP} and\cite{T.G.A} for three and four
dimensional Lie algebras respectively). On the other hand, the
automorphism groups and ad-invariant metrics can be used in the
calculation of complex and bi-Hermitian structures \cite{RS} and
generalized complex structures \cite{SRS} on Lie algebras and
also in the classification of Lie bialgebras \cite{ERF}.
Meanwhile, the calculation of ad-invariant metric on Lie algebras
are important in the construction of the physical models such as
WZW models \cite{NW}. Here in this manner we calculate the
automorphism group and ad-invariant metric on six dimensional
solvable Lie algebras (with 5 \cite{M} and 4 \cite{T} and 3
(nilpotent) \cite{MO} dimensional nilradicals).
%%%%%%%%%%%%%%%%%%%%%%%%%%%%%%%%%%%%%%%%%%%%%%%%%%%%%%%%%%%%%%%%%%%%%%%%%%%%%%%%%%%%%%%%%%%
\section{\bf Mathematical preliminaries}

Let L be Lie algebra with the base $\{X_{i}\}$; then we have
\begin{equation}
[X_{i},X_{j}]=f_{ij}\hspace{0cm}^{k}X_{k},
\end{equation}
where $f_{ij}\hspace{0cm}^{k}$ are the structure constants of the
Lie algebra L. An automorphism O is a linear map on L such that
preserves the Lie algebra structure i.e.:
\begin{equation}
O [X_{i},X_{j}]=[O X_{i},O X_{j}].
\end{equation}
Or, by use of $O X_{i}=X'_{i}$ we have
\begin{equation}
[X'_{i},X'_{j}]=f_{ij}\hspace{0cm}^{k}X'_{k},
\end{equation}
i.e. the automorphism O is a linear map which fixes the structure
constants. By use of matrix representation for O ; i.e.
$OX_{i}=O_{i}\hspace{0cm}^{j} X_{j}$ we have rewritten relation
$(3)$ in the following form;
\begin{equation}
O_{l}\hspace{0cm}^{i} O_{m}\hspace{0cm}^{j}
f_{ij}\hspace{0cm}^{k}= f_{lm}\hspace{0cm}^{n}
O_{n}\hspace{0cm}^{k},
\end{equation}
where by use of adjoint representation
$(\chi_{i})_{j}\hspace{0cm}^{k}=-f_{ij}\hspace{0cm}^{k}$ or
$({\cal{Y}}^{k})_{ij}=-f_{ij}\hspace{0cm}^{k}$ one can rewrite
relation $(4)$ in the following matrix form:
\begin{equation}
O_{l}\hspace{0cm}^{i} O \chi_{i}=\chi_{l} O,
\end{equation}
or
\begin{equation}
O {\cal{Y}}^{k} O^{t}={\cal{Y}}^{n}O_{n}\hspace{0cm}^{k}.
\end{equation}
In this way, by use of the above relations one can calculate the
automorphism group O of a Lie algebra L. Furthermore, an
ad-invariant symmetric metric{\footnote{The Cartan-Killing form
$K_{ij}=f_{ik}\hspace{0cm}^{l}f_{jl}\hspace{0cm}^{k}$is a special
case of this metric.}} on Lie algebra L can be written as follow:
\begin{equation}
\langle X_{i}, X_{j}\rangle=g_{ij},
\end{equation}
such that
\begin{equation}
ad_{X_{j}}\langle X_{i}, X_{k}\rangle=0,
\end{equation}
i.e.
\begin{equation}
\langle ad_{X_{j}}X_{i},X_{k}\rangle + \langle
X_{i},ad_{X_{j}}X_{k}\rangle = 0,
\end{equation}
or
\begin{equation}
\langle X_{i},[X_{j},X_{k}]\rangle=\langle[X_{i},X_{j}],X_{k}\rangle,
\end{equation}
which can be rewritten in the following matrix form:
\begin{equation}
\chi_{i}g=-(\chi_{i}g)^{t}.
\end{equation}
Note that because the automorphism map O fixes the structure
constants (see $(3)$); from above relation $(11)$ one can see that
the metric g on Lie algebra L is also fixed under automorphism
map, i.e:
\begin{equation}
g'_{ij}=\langle X_{i},X_{j}\rangle = \langle
X'_{i},X'_{j}\rangle=g_{ij},
\end{equation}
or
\begin{equation}
g_{ij}=O_{i}\hspace{0cm}^{k} O_{j}\hspace{0cm}^{l} g_{kl},
\end{equation}
with matrix form{\footnote{This relation can also be obtain by
replacing $(5)$ into $(11)$ directly.}}
\begin{equation}
g = O g O^{t}.
\end{equation}
i.e. the ad-invariant metric is also invariant under automorphism
group or in other word the automorphism groups are isometries of
this metric. Now, one can calculate ad-invariant metric g on Lie
algebras according to the relations $(11)$ and $(14)$. In the
next section we calculate the automorphism groups and
ad-invariant metric on six dimensional solvable real Lie algebras
by use of maple program for solving relations $(5)$,$(11)$ and
$(14)$.
%%%%%%%%%%%%%%%%%%%%%%%%%%%%%%%%%%%%%%%%%%%%%%%%%%%%%%%%%%%%%%%%%%%%%%%%
\section{\bf Automorphism groups and ad-invariant metrics}

Classification of six dimensional solvable real Lie algebras with
3 dimensional nilradical (i.e. six dimensional nilpotent Lie
algebras) are obtained by Morozov \cite{MO} then Lie algebras
with 5 dimensional nilradical are mainly classified by
Mubarakzyanov \cite{M} and were finished by Turkowski with the
classification of these Lie algebras with 4 dimensional
nilradical \cite{T}; (for a good bibilography see \cite{PT}).
Here we use the Mubarakzyanov \cite{M} classification of six
dimensional solvable real Lie algebras with nilradical 5 (see
also \cite{C} for correction of some misprint of \cite{M}). The
automorphism groups{\footnote{Note that we choose automorphism
groups, (i.e. matrices that connected to the identity matrix)
from the solutions of the relation 5 or 6.}} and structure
constants of these 99 Lie algebras are written in table 1. For 40
six dimensional solvable real Lie algebras with nilradical 4
\cite{T}, these are written in table 2. Tables 3 contains the
automorphism groups for 22 six dimensional nilpotent Lie algebras
\cite{MO}(and also use \cite{PSW}). The ad-invariant metrics of
all six dimensional real solvable Lie algebras are written in table 4.\\
%%%%%%%%%%%%%%%%%%%%%%%%%%%%%%%%%%%%%%%%%%%%%%%%%%%%%%%%%%%%%%%%%%%%%%%%%%%%%%%%%%%%%%%%%%%%%%%%%%%%%%%%%%%%
\newpage
\landscape
 \vspace{8mm}
\begin{center}
\begin{tabular}{c c c}
\multicolumn{3}{c}{TABLE 1: The automorphism groups of six dimensional solvable real Lie algebras with nilradical 5 }\\
  \hline
  \hline
Lie Algebra & Non Vanishing & Automorphism group \\&Structure Constants&\\
\hline \hline \vspace{-3.5mm}
$g_{6,1}$\\
$0<\mid\delta\mid\leq\mid\gamma\mid\leq\mid\beta\mid\leq\mid\alpha\mid\leq1$
&$f^{1}_{16}=1,f^{2}_{26}=\alpha,f^{3}_{36}=\beta
,f^{4}_{46}=\gamma,f^{5}_{56}=\delta$ &
$\left(\begin{array}{cccccc}
a_{1} & 0 & 0 & 0 & 0 & 0 \\
0 & a_{2} & 0 & 0 & 0 & 0 \\
0 & 0 & a_{3} & 0 & 0 & 0 \\
0 & 0 & 0 & a_{4}& 0 & 0 \\
0 & 0 & 0& 0 & a_{5} & 0\\
a_{6} & a_{7} & a_{8}& a_{9}& a_{10} & 1\\
\end{array}
\right)$\\
%%%%%%%%%%%%%%%%%%%%%%%%%%%%%%
\vspace{-3.5mm}
$g_{6,2}$\\
$0<\mid\delta\mid\leq\mid\gamma\mid\leq1$
&$f^{1}_{16}=\alpha,f^{1}_{26}=1,f^{2}_{26}=\alpha,f^{3}_{36}=1,f^{4}_{46}=\gamma,f^{5}_{56}=\delta$
& $\left(\begin{array}{cccccc}
a_{2} & 0 & 0 & 0 & 0 & 0 \\
a_{1} & a_{2} & 0 & 0 & 0 & 0 \\
0 & 0 & a_{3} & 0 & 0 & 0 \\
0 & 0 & 0 & a_{4}& 0 & 0 \\
0 & 0 & 0& 0 & a_{5} & 0\\
a_{6} & a_{7} & a_{8}& a_{9}& a_{10} & 1\\
\end{array}
\right)$  \\
%%%%%%%%%%%%%%%%%%%%%%%%%%%%%%%
\vspace{-3.5mm}
$g_{6,3}$\\
\end{tabular}
\end{center}
%%%%%%%%%%%%%%%%%%%%%%%%%%%%%%%%%%%%%%%%%%%%%%%%%%%%%%%%%%%%%%%%%%%%%%%%%%%%%%%%%%%%%%%%%%%%%%%%%%%%%%%%%%%%%%%%%%%%%%%%%%%%
\newpage
\vspace{10mm}
\begin{center}
\begin{tabular}{c c c}
\multicolumn{3}{c}{TABLE 1. (Continued) }\\
  \hline
  \hline
Lie Algebra & Non Vanishing & Automorphism group \\&Structure Constants&\\
\hline \hline
%%%%%%%%%%%%%%%%%%%%%%%%%%%%%%%
\vspace{-3.5mm}
$g_{6,6}$\\
\vspace{-3.5mm} $\alpha\leq\beta$
&$f^{1}_{16}=1,f^{2}_{26}=\alpha,f^{2}_{36}=1,f^{3}_{36}=\alpha,f^{4}_{46}=\beta$
& $\left(\begin{array}{cccccc}
a_{1} & 0 & 0 & 0 & 0 & 0 \\
0 & a_{3} & 0 & 0 & 0 & 0 \\
0 & a_{2}& a_{3} & 0 & 0 & 0 \\
0 & 0 & 0 & a_{5}& 0 & 0 \\
0 & 0 & 0& a_{4}& a_{5} & 0\\
a_{6} & a_{7} & a_{8}& a_{9}& a_{10} & 1\\
\end{array}
\right)$ \\
& $,f^{4}_{56}=1,f^{5}_{56}=\beta $\\
%%%%%%%%%%%%%%%%%%%%%%%%%%%%%%%%%
\vspace{-3.5mm}
$g_{6,7}$\\
\vspace{-3.5mm} $\alpha^{2}+\beta^{2}\neq0$
&$f^{1}_{16}=\alpha,f^{1}_{26}=1,f^{2}_{26}=\alpha,f^{2}_{36}=1,f^{3}_{36}=\alpha$
& $\left(\begin{array}{cccccc}
a_{3} & 0 & 0 & 0 & 0 & 0 \\
a_{2} & a_{3} & 0 & 0 & 0 & 0 \\
a_{1} & a_{2}& a_{3} & 0 & 0 & 0 \\
0 & 0 & 0 & a_{5}& 0 & 0 \\
0 & 0 & 0& a_{4} & a_{5} & 0\\
a_{6} & a_{7} & a_{8}& a_{9}& a_{10} & 1 \\
\end{array}
\right)$ \\
\end{tabular}
\end{center}
%%%%%%%%%%%%%%%%%%%%%%%%%%%%%%%%%%%%%%%%%%%%%%%%%%%%%%%%%%%%%%%%%%%%%%%%%%%%%%%%%%%%%%%%%%%%%%%%%%%%%%%%%%%%%%%%%%%%%%%%%%%
\newpage
\vspace{10mm}
\begin{center}
\begin{tabular}{c c c}
\multicolumn{3}{c}{TABLE 1. (Continued)}\\
\hline
\hline
Lie Algebra & Non Vanishing & Automorphism group \\&Structure Constants&\\
\hline
\hline
%%%%%%%%%%%%%%%%%%%%%%%%%%%%%%%%%%%%
\vspace{-3.5mm}
$g_{6,12}$\\
\vspace{-3.5mm}
 $\alpha\neq0$
&$f^{1}_{16}=\alpha,f^{2}_{26}=p,f^{3}_{26}=-1,f^{2}_{36}=1,f^{3}_{36}=p,f^{2}_{46}=1$
& $\left(\begin{array}{cccccc}
a_{1} & 0 & 0 & 0 & 0 & 0 \\
0 & a_{4} & a_{5} & 0 & 0 & 0 \\
0 & -a_{5} & a_{4} & 0 & 0 & 0 \\
0 & a_{2}& a_{3} & a_{4}& a_{5} & 0 \\
0 & -a_{3}& a_{2} & -a_{5}& a_{4} & 0\\
a_{6} & a_{7} & a_{8}& a_{9}& a_{10} & 1\\
\end{array}
\right)$ \\
& $,f^{4}_{46}=p,f^{5}_{46}=-1,f^{3}_{56}=1,f^{4}_{56}=1 ,f^{5}_{56}=p$\\
%%%%%%%%%%%%%%%%%%%%%%%%%%%%%%%%%%%
\vspace{-3.5mm}
$g_{6,13}$\\
\vspace{-3.5mm} \small{if  $\alpha^{2}+\beta^{2}=0$ then $h=0$ }
&$f^{1}_{23}=1,f^{1}_{16}=\alpha+\beta,f^{2}_{26}=\alpha,f^{3}_{36}=\beta$
& $\left(\begin{array}{cccccc}
a_{2}a_{4} & 0 & 0 & 0 & 0 & 0 \\
a_{1} & a_{2} & 0 & 0 & 0 & 0 \\
a_{3} & 0 & a_{4} & 0 & 0 & 0 \\
0 & 0 & 0 & a_{5}& 0 & 0 \\
0 & 0 & 0& 0 & a_{6} & 0\\
a_{7} & \frac{a_{3}\alpha}{a_{4}} & -\frac{a_{1}\beta}{a_{2}}& a_{8}& a_{9} & 1\\
\end{array}
\right)$ \\
\end{tabular}
\end{center}
%%%%%%%%%%%%%%%%%%%%%%%%%%%%%%%%%%%%%%%%%%%%%%%%%%%%%%%%%%%%%%%%%%%%%%%%%%%%%%%%%%%%%%%%
\newpage
\vspace{10mm}
\begin{center}
\begin{tabular}{c c c}
\multicolumn{3}{c}{TABLE 1. (Continued)}\\
\hline \hline
Lie Algebra & Non Vanishing & Automorphism group \\&Structure Constants&\\
\hline \hline
%%%%%%%%%%%%%%%%%%%%%%%%%%%%%%%%%%%%
\vspace{-3.5mm}
$g_{6,18}$\\
\vspace{-3.5mm} $\beta\neq0$
&$f^{1}_{23}=1,f^{1}_{16}=1+\alpha,f^{2}_{26}=\alpha,f^{3}_{36}=1,f^{4}_{36}=1$
& $\left(\begin{array}{cccccc}
a_{1}a_{2} & 0 & 0 & 0 & 0 & 0 \\
-a_{1}a_{7} & a_{1} & 0 & 0 & 0 & 0 \\
\frac{a_{2}a_{6}}{\alpha} & 0 & a_{2} & a_{3} & 0 & 0 \\
0 & 0 & 0 & a_{2}& 0 & 0 \\
0 & 0 & 0 & 0 & a_{4} & 0\\
a_{5} & a_{6} & a_{7}& a_{8}& a_{9} & 1\\
\end{array}
\right)$ \\
& $,f^{4}_{46}=1,f^{5}_{56}=\beta$\\
%%%%%%%%%%%%%%%%%%%%%%%%%%%%%%%%%%%
\vspace{-3.5mm}
$g_{6,19}$\\
\vspace{-3.5mm}
&$f^{1}_{23}=1,f^{1}_{16}=1+\alpha,f^{2}_{26}=\alpha,f^{3}_{36}=1,f^{4}_{36}=1$
& $\left(\begin{array}{cccccc}
a_{1}a_{2} & 0 & 0 & 0 & 0 & 0 \\
-a_{1}a_{7} & a_{1} & 0 & 0 & 0 & 0 \\
\frac{a_{2}a_{6}}{\alpha} & 0 & a_{2} & a_{3} & 0 & 0 \\
0 & 0 & 0 & a_{2}& 0 & 0 \\
a_{4} & 0 & 0 & 0 & a_{1}a_{2} & 0\\
a_{5} & a_{6} & a_{7}& a_{8}& a_{9} & 1\\
\end{array}
\right)$ \\
\end{tabular}
\end{center}
%%%%%%%%%%%%%%%%%%%%%%%%%%%%%%%%%%%%%%%%%%%%%%%%%%%%%%%%%%%%%%%%%%%%%%%%%%%%%%%%%%%%%%%%%%%%%%%%%%%%%%%%%%%%%%%%%%%%%%%%%%%
\newpage
\vspace{10mm}
\begin{center}
\begin{tabular}{c c c}
\multicolumn{3}{c}{TABLE 1. (Continued)}\\
\hline \hline
Lie Algebra & Non Vanishing & Automorphism group \\&Structure Constants&\\
\hline \hline
%%%%%%%%%%%%%%%%%%%%%%%%%%%%%%%%%%%%
\vspace{-3.5mm}
$g_{6,24}$\\
&$f^{1}_{23}=1,f^{3}_{26}=1,f^{4}_{36}=1,f^{1}_{46}=h,f^{5}_{56}=1$
& $\left(\begin{array}{cccccc}
1 & 0 & 0 & 0 & 0 & 0 \\
a_{1} & 1 & a_{2} & a_{3} & 0 & 0 \\
ha_{3}+a_{7}-a_{2}a_{6} & 0 & 1 & a_{2} & 0 & 0 \\
ha_{2}-a_{6} & 0 & 0 & 1 & 0 & 0 \\
0 & 0 & 0& 0 & a_{4} & 0\\
a_{5} & a_{6} & a_{7}& a_{8}& a_{9} & 1\\
\end{array}
\right)$ \\
%%%%%%%%%%%%%%%%%%%%%%%%%%%%%%%%%%%
\vspace{-3.5mm}
$g_{6,25}$\\
\vspace{-3.5mm}
&$f^{1}_{23}=1,f^{1}_{16}=1+h,f^{2}_{26}=1,f^{3}_{36}=h$ &
$\left(\begin{array}{cccccc}
a_{1}a_{2} & 0 & 0 & 0 & 0 & 0 \\
-\frac{a_{1}a_{7}}{h} & a_{1} & 0 & 0 & 0 & 0 \\
a_{2}a_{6}& 0 & a_{2} & 0 & 0 & 0 \\
0 & 0 & 0 & a_{3} & a_{4} & 0\\
0 & 0 & 0 & 0 & a_{3} & 0 \\
a_{5} & a_{6} & a_{7} & a_{8}& a_{9} & 1\\
\end{array}
\right)$ \\
 &$,f^{4}_{46}=\beta,f^{5}_{46}=1,f^{5}_{56}=\beta $\\
\end{tabular}
\end{center}
%%%%%%%%%%%%%%%%%%%%%%%%%%%%%%%%%%%%%%%%%%%%%%%%%%%%%%%%%%%%%%%%%%%%%%%%%%%%%%%%%%%%%%%%
\newpage
\vspace{10mm}
\begin{center}
\begin{tabular}{c c c}
\multicolumn{3}{c}{TABLE 1. (Continued)}\\
\hline \hline
Lie Algebra & Non Vanishing & Automorphism group \\&Structure Constants&\\
\hline \hline
%%%%%%%%%%%%%%%%%%%%%%%%%%%%%%%%%%%%
\vspace{-3.5mm}
$g_{6,30}$\\
\vspace{-3.5mm} &$f^{1}_{23}=1,f^{3}_{26}=1,f^{4}_{46}=1$ &
$\left(\begin{array}{cccccc}
a_{2}^{2} & 0 & 0 & 0 & 0 & 0 \\
a_{1}& a_{2} & a_{3} & 0 & 0 & 0 \\
a_{2}a_{7} & 0 & a_{2} & 0 & 0 & 0\\
0 & 0 & 0 & a_{4} & a_{5} & 0 \\
0 & 0 & 0 & 0 & a_{4} & 0 \\
a_{6} & 0 & a_{7} & a_{8}& a_{9} & 1\\
\end{array}
\right)$ \\
& $,f^{5}_{46}=1,f^{5}_{56}=1$\\
%%%%%%%%%%%%%%%%%%%%%%%%%%%%%%%%%%%
\vspace{-3.5mm}
$g_{6,31}$\\
\vspace{-3.5mm}
&$f^{1}_{23}=1,f^{1}_{16}=2,f^{2}_{26}=1,f^{3}_{26}=1,f^{3}_{36}=1$
& $\left(\begin{array}{cccccc}
a_{1}^{2} & 0 & 0 & 0 & 0 & 0 \\
a_{1}a_{6}+ a_{2} a_{6}- a_{1} a_{7}& a_{1} & a_{2} & a_{3} & a_{4} & 0 \\
a_{1}a_{6} & 0 & a_{1} & a_{2} & a_{3} & 0\\
0 & 0 & 0 & a_{1} & a_{2} & 0 \\
0 & 0 & 0 & 0 & a_{1} & 0 \\
a_{5} & a_{6} & a_{7} & a_{8}& a_{9} & 1\\
\end{array}
\right)$ \\
\end{tabular}
\end{center}
%%%%%%%%%%%%%%%%%%%%%%%%%%%%%%%%%%%%%%%%%%%%%%%%%%%%%%%%%%%%%%%%%%%%%%%%%%%%%%%%%%%%%%%%
\newpage
\vspace{10mm}
\begin{center}
\begin{tabular}{c c c}
\multicolumn{3}{c}{TABLE 1. (Continued) }\\
\hline \hline
Lie Algebra & Non Vanishing & Automorphism group \\&Structure Constants&\\
\hline \hline
%%%%%%%%%%%%%%%%%%%%%%%%%%%%%%%%%%%%
\vspace{-3.5mm}
$g_{6,36}$\\
\vspace{-3.5mm}
&$f^{1}_{23}=1,f^{1}_{16}=2\alpha,f^{2}_{26}=\alpha,f^{3}_{26}=1,f^{3}_{36}=\alpha$
& $\left(\begin{array}{cccccc}
a_{2}^{2} & 0 & 0 & 0 & 0 & 0 \\
a_{1}& a_{2} & a_{3} & 0 & 0 & 0 \\
\frac{ a_{2}a_{7}}{\alpha} & 0 & a_{2} & 0 & 0 & 0\\
0 & 0 & 0 & a_{4} &  a_{5} & 0 \\
0 & 0 & 0 & -a_{5} & a_{4} & 0 \\
a_{6} & a_{7} &\frac{a_{2}a_{7}+\alpha a_{3} a_{7}-\alpha^{2}a_{1}}{a_{2}\alpha} & a_{8}& a_{9} & 1\\
\end{array}
\right)$ \\
& $,f^{4}_{46}=\gamma,f^{5}_{46}=1,f^{4}_{56}=-1,f^{5}_{56}=\gamma$\\
%%%%%%%%%%%%%%%%%%%%%%%%%%%%%%%%%%%
\vspace{-3.5mm}
$g_{6,37}$\\
\vspace{-3.5mm} $s\neq0$
&$f^{1}_{23}=1,f^{1}_{16}=2\alpha,f^{2}_{26}=\alpha,f^{3}_{26}=1,f^{2}_{36}=-1$
& $\left(\begin{array}{cccccc}
a_{1}^{2}+a_{2}^{2} & 0 & 0 & 0 & 0 & 0 \\
\frac{a_{1}a_{6}+a_{2}a_{7}+\alpha a_{2}a_{6}-\alpha a_{1}a_{7}}{1+\alpha^{2}}& a_{1} & a_{2} & 0 & 0 & 0 \\
\frac{a_{1}a_{7}- a_{2}a_{6}+\alpha a_{1}a_{6}+\alpha a_{2}a_{7}}{1+\alpha^{2}}& -a_{2} & a_{1} & 0 & 0 & 0 \\
0 & 0 & 0 & a_{4} & -a_{3} & 0\\
0 & 0 & 0 & a_{3} & a_{4} & 0 \\
a_{5} & a_{6} & a_{7} & a_{8}& a_{9} & 1\\
\end{array}
\right)$ \\
\end{tabular}
\end{center}
%%%%%%%%%%%%%%%%%%%%%%%%%%%%%%%%%%%%%%%%%%%%%%%%%%%%%%%%%%%%%%%%%%%%%%%%%%%%%%%%%%%%%%%%
\newpage
\vspace{10mm}
\begin{center}
\begin{tabular}{c c c}
\multicolumn{3}{c}{TABLE 1. (Continued)}\\
  \hline
  \hline
Lie Algebra & Non Vanishing & Automorphism group \\&Structure Constants&\\
\hline \hline
%%%%%%%%%%%%%%%%%%%%%%%%%%%%%%%%%%%%
\vspace{-3.5mm}
$g_{6,42}$\\
\vspace{-3.5mm}
&$f^{2}_{15}=1,f^{1}_{45}=1,f^{1}_{16}=1+h,f^{2}_{26}=h+2$ &
$\left(\begin{array}{cccccc}
a_{1}a_{5} & -a_{1}a_{5}a_{8} & 0 & 0 & 0 & 0 \\
0 & a_{1}a_{5}^{2} & 0 & 0 & 0 & 0 \\
0 & 0 & a_{5} & 0 & 0 & 0 \\
-a_{1}a_{8} & \frac{a_{1}a_{8}^{2}}{2} & 0 & a_{1} & 0 & 0 \\
a_{2} & a_{3} & a_{4}& 0 & a_{5} & 0\\
\frac{a_{2}a_{8}+ha_{3}+a_{3}}{a_{5}} & a_{6} & a_{7}& \frac{ha_{2}}{a_{5}}& a_{8} & 1\\
\end{array}
\right)$ \\
& $,f^{3}_{36}=1,f^{4}_{46}=h,f^{3}_{56}=1,f^{5}_{56}=1$\\
%%%%%%%%%%%%%%%%%%%%%%%%%%%%%%%%%%%
\vspace{-3.5mm}
$g_{6,43}$\\
\vspace{-3.5mm}
&$f^{2}_{15}=1,f^{1}_{45}=1,f^{2}_{26}=1,f^{2}_{36}=1$ &
$\left(\begin{array}{cccccc}
 1 & -a_{8} & 0 & 0 & 0 & 0 \\
 0 & a_{3} & 0 & 0 & 0 & 0 \\
 0& a_{2}-a_{3}a_{4}-a_{3}a_{7}a_{8} & a_{3} & 0 & 0 & 0 \\
-\frac{a_{8}}{a_{3}} & \frac{a_{8}^{2}}{2a_{3}} & 0 & \frac{1}{a_{3}}& 0 & 0 \\
-a_{3}a_{7} & a_{1} & a_{2} & 0 & a_{3} & 0\\
a_{4} &  a_{5} &  a_{6}& a_{7}& a_{8} & 1\\
\end{array}
\right)$ \\
\end{tabular}
\end{center}
%%%%%%%%%%%%%%%%%%%%%%%%%%%%%%%%%%%%%%%%%%%%%%%%%%%%%%%%%%%%%%%%%%%%%%%%%%%%%%%%%%%%%%%%
\newpage
\vspace{10mm}
\begin{center}
\begin{tabular}{c c c}
\multicolumn{3}{c}{TABLE 1. (Continued)}\\
  \hline
  \hline
Lie Algebra & Non Vanishing & Automorphism group \\&Structure Constants&\\
\hline \hline
%%%%%%%%%%%%%%%%%%%%%%%%%%%%%%%%%%%%
\vspace{-3.5mm}
$g_{6,47}$\\
\vspace{-3.5mm} $\gamma\neq 0$&
$f^{2}_{15}=1,f^{1}_{45}=1,f^{1}_{16}=1,f^{2}_{26}=1$ &
$\left(\begin{array}{cccccc}
a_{4}a_{5} & a_{2}a_{5} & 0 & 0 & 0 & 0 \\
0 & a_{4}a_{5}^{2} & 0 & 0 & 0 & 0 \\
0 & 0 & a_{1} & 0 & 0 & 0\\
a_{2} & a_{3} & 0 & a_{4} & 0 & 0 \\
a_{5}a_{9} & a_{5}a_{6} & 0 & 0 & a_{5} & 0 \\
a_{6} & a_{7} & a_{8} & a_{9}& 0 & 1\\
\end{array}
\right)$ \\
$\epsilon=0$& $,f^{3}_{36}=\gamma,f^{4}_{46}=1$\\
%%%%%%%%%%%%%%%%%%%%%%%%%%%%%%%%%%%%
\vspace{-3.5mm}
$g_{6,48}$\\
\vspace{-3.5mm}
&$f^{2}_{15}=1,f^{1}_{45}=1,f^{1}_{16}=1,f^{2}_{26}=1$ &
$\left(\begin{array}{cccccc}
a_{4}a_{5} & a_{2}a_{5} & 0 & 0 & 0 & 0 \\
0 & a_{4}a_{5}^{2} & 0 & 0 & 0 & 0 \\
0& a_{1} & a_{4}a_{5}^{2} & 0 & 0 & 0 \\
a_{2} & a_{3} & 0 & a_{4} & 0 & 0\\
a_{5}a_{9} & a_{5}a_{6} & 0 & 0 & a_{5} & 0 \\
a_{6} & a_{7} & a_{8} & a_{9}& 0 & 1\\
\end{array}
\right)$ \\
\end{tabular}
\end{center}
%%%%%%%%%%%%%%%%%%%%%%%%%%%%%%%%%%%%%%%%%%%%%%%%%%%%%%%%%%%%%%%%%%%%%%%%%%%%%%%%%%%%%%%%
\newpage
\vspace{10mm}
\begin{center}
\begin{tabular}{c c c}
\multicolumn{3}{c}{TABLE 1. (Continued)}\\
  \hline
  \hline
Lie Algebra & Non Vanishing & Automorphism group \\&Structure Constants&\\
\hline \hline
%%%%%%%%%%%%%%%%%%%%%%%%%%%%%%%%%%%
\vspace{-3.5mm}
$g_{6,50}$\\
\vspace{-3.5mm} $\epsilon=-1$
&$f^{2}_{15}=1,f^{1}_{45}=1,f^{1}_{16}=1,f^{3}_{36}=1$ &
$\left(\begin{array}{cccccc}
a_{4} & a_{1} & 0 & 0 & 0 & 0 \\
0 & a_{4} & 0 & 0 & 0 & 0 \\
0 & -a_{3} & a_{4} & 0 & 0 & 0 \\
a_{1} & a_{2} & a_{3} & a_{4} & 0 & 0\\
a_{8} & a_{5} & 0 & 0 & 1 & 0 \\
a_{5} & a_{6} & a_{7}& a_{8}& 0 & 1\\
\end{array}
\right)$ \\
& $,f^{2}_{26}=1,f^{2}_{36}=-1,f^{3}_{46}=1,f^{4}_{46}=1$\\
%%%%%%%%%%%%%%%%%%%%%%%%%%%%%%%%%%%
\vspace{-3.5mm}
$g_{6,51}$\\
\vspace{-3.5mm}
 $\epsilon=1$ &$f^{2}_{15}=1,f^{1}_{45}=1$ &
$\left(\begin{array}{cccccc}
a_{4} & a_{2} & 0 & 0 & 0 & 0 \\
0 & a_{4} & 0 & 0 & 0 & 0 \\
0 & 0 & a_{1} & 0 & 0 & 0 \\
a_{2} & a_{3} & 0 & a_{4} & 0 & 0\\
a_{5} & a_{6} & 0 & a_{7} & 1 & 0 \\
a_{7} & a_{8} & a_{9} & 0 & 0 & 1\\
\end{array}
\right)$ \\
\end{tabular}
\end{center}
%%%%%%%%%%%%%%%%%%%%%%%%%%%%%%%%%%%%%%%%%%%%%%%%%%%%%%%%%%%%%%%%%%%%%%%%%%%%%%%%%%%%%%%%%
\newpage
\vspace{10mm}
\begin{center}
\begin{tabular}{c c c}
\multicolumn{3}{c}{TABLE 1. (Continued)}\\
  \hline
  \hline
Lie Algebra & Non Vanishing & Automorphism group \\&Structure Constants&\\
\hline \hline
%%%%%%%%%%%%%%%%%%%%%%%%%%%%%%%%%%%
\vspace{-3.5mm}
$g_{6,53}$\\
\vspace{-3.5mm} &$f^{1}_{35}=1,f^{2}_{45}=1,f^{3}_{36}=1$ &
$\left(\begin{array}{cccccc}
a_{1}a_{5} & a_{2}a_{5} & 0 & 0 & 0 & 0 \\
a_{3}a_{5} & a_{4}a_{5} & 0 & 0 & 0 & 0 \\
a_{1}a_{10}& a_{2}a_{10} & a_{1} & a_{2} & 0 & 0 \\
a_{3}a_{10} & a_{4}a_{10} & a_{3} & a_{4}& 0 & 0 \\
a_{5}a_{8} & a_{5}a_{9} & 0 & 0 & a_{5} & 0\\
a_{6} & a_{7} & a_{8}& a_{9}& a_{10} & 1\\
\end{array}
\right)$ \\
& $,f^{4}_{46}=1,f^{5}_{56}=-1$\\
%%%%%%%%%%%%%%%%%%%%%%%%%%%%%%%%%%%
\vspace{-3.5mm}
$g_{6,54}$\\
\vspace{-3.5mm}
&$f^{1}_{35}=1,f^{2}_{45}=1,f^{1}_{16}=1,f^{2}_{26}=\lambda$ &
$\left(\begin{array}{cccccc}
a_{1}a_{6} & 0 & 0 & 0 & 0 & 0 \\
0 & a_{3}a_{6} & 0 & 0 & 0 & 0 \\
\frac{a_{1}a_{2}}{a_{3}} & 0 & a_{1} & 0 & 0 & 0\\
0 & a_{2} & 0 & a_{3} & 0 & 0 \\
a_{4} & a_{5} & 0 & 0 & a_{6} & 0 \\
a_{7} & a_{8} & \frac{a_{4}(1-\gamma)}{a_{6}} & \frac{a_{5}(\lambda -\gamma)}{a_{6}}& -\frac{a_{2}\gamma}{a_{3}} & 1\\
\end{array}
\right)$ \\
\end{tabular}
\end{center}
%%%%%%%%%%%%%%%%%%%%%%%%%%%%%%%%%%%%%%%%%%%%%%%%%%%%%%%%%%%%%%%%%%%%%%%%%%%%%%%%%%%%%%%%
\newpage
\vspace{10mm}
\begin{center}
\begin{tabular}{c c c}
\multicolumn{3}{c}{TABLE 1. (Continued)}\\
  \hline
  \hline
Lie Algebra & Non Vanishing & Automorphism group \\&Structure Constants&\\
\hline \hline
%%%%%%%%%%%%%%%%%%%%%%%%%%%%%%%%%%%
\vspace{-3.5mm}
$g_{6,58}$\\
\vspace{-3.5mm} $\omega =0$
&$f^{1}_{35}=1,f^{2}_{45}=1,f^{1}_{16}=3,f^{2}_{26}=2,f^{2}_{36}=1$
& $\left(\begin{array}{cccccc}
a_{2}a_{4}^{2} & 0 & 0 & 0 & 0 & 0 \\
0 & a_{2}a_{4} & 0 & 0 & 0 & 0 \\
-a_{2}a_{4}a_{9} & a_{1} & a_{2}a_{4} & 0 & 0 & 0\\
0 & -a_{2}a_{9} & 0 & a_{2} & 0 & 0 \\
\frac{a_{4}a_{7}}{2} & a_{4}a_{8}-a_{3}a_{9} & 0 & a_{3} & a_{4} & 0\\
a_{5} & a_{6} & a_{7} & a_{8} & a_{9} & 1 \\
\end{array}
\right)$ \\
& $,f^{3}_{36}=2,f^{4}_{46}=1,f^{5}_{56}=1$\\
%%%%%%%%%%%%%%%%%%%%%%%%%%%%%%%%%%%
\vspace{-3.5mm}
$g_{6,59}$\\
\vspace{-3.5mm}
&$f^{1}_{35}=1,f^{2}_{45}=1,f^{1}_{16}=1,f^{3}_{36}=1$ &
$\left(\begin{array}{cccccc}
a_{2} & 0 & 0 & 0 & 0 & 0 \\
0 & 1 & 0 & 0 & 0 & 0 \\
a_{1} & 0 & a_{2} & 0 & 0 & 0\\
0 & a_{3} & 0 & 1 & 0 & 0 \\
a_{7} & a_{4} & 0 & ha_{3}+a_{8} & 1 & 0\\
a_{5} & a_{6} & a_{7} & a_{8} & 0 & 1 \\
\end{array}
\right)$ \\
\end{tabular}
\end{center}
%%%%%%%%%%%%%%%%%%%%%%%%%%%%%%%%%%%%%%%%%%%%%%%%%%%%%%%%%%%%%%%%%%%%%%%%%%%%%%%%%%%%%%%%
\newpage
\vspace{10mm}
\begin{center}
\begin{tabular}{c c c}
\multicolumn{3}{c}{TABLE 1. (Continued)}\\
  \hline
  \hline
Lie Algebra & Non Vanishing & Automorphism group \\&Structure Constants&\\
\hline
\hline
%%%%%%%%%%%%%%%%%%%%%%%%%%%%%%%%%%%%%%%%%%%%%%%%%%%%%%%%%%%%%%%%%%%%%%%%%%%%%%%%%%%%%%%
\vspace{-3.5mm}
$g_{6,62}$\\
\vspace{-3.5mm} $\omega = 0$
&$f^{1}_{35}=1,f^{2}_{45}=1,f^{1}_{16}=2,f^{2}_{26}=1$ &
$\left(\begin{array}{cccccc}
a_{2}a_{4}^{2} & 0 & 0 & 0 & 0 & 0 \\
0 & a_{2}a_{4} & 0 & 0 & 0 & 0 \\
-a_{2}a_{4}a_{9} & a_{1} & a_{2}a_{4} & 0 & 0 & 0\\
0 & -a_{2}a_{9} & 0 & a_{2} & 0 & 0 \\
a_{4}a_{7}-a_{4}a_{8}a_{9} & a_{3} & a_{4}a_{8} & 0 & a_{4} & 0 \\
a_{5} & a_{6} & a_{7} & a_{8}& a_{9} & 1\\
\end{array}
\right)$ \\
& $,f^{2}_{36}=1,f^{3}_{36}=1,f^{5}_{56}=1$\\
%%%%%%%%%%%%%%%%%%%%%%%%%%%%%%%%%%%
\vspace{-3.5mm}
$g_{6,63}$\\
\vspace{-3.5mm}
&$f^{1}_{35}=1,f^{2}_{45}=1,f^{1}_{16}=1,f^{2}_{26}=\lambda$ &
$\left(\begin{array}{cccccc}
a_{2} & 0 & 0 & 0 & 0 & 0 \\
0 & a_{4} & 0 & 0 & 0 & 0 \\
a_{1} & 0 & a_{2} & 0 & 0 & 0 \\
0 & a_{3} & 0 & a_{4} & 0 & 0\\
a_{8} & a_{5} & 0 & 0 & 1 & 0 \\
a_{6} & a_{7} & a_{8}& \lambda a_{5}& 0 & 1\\
\end{array}
\right)$ \\
\end{tabular}
\end{center}
%%%%%%%%%%%%%%%%%%%%%%%%%%%%%%%%%%%%%%%%%%%%%%%%%%%%%%%%%%%%%%%%%%%%%%%%%%%%%%%%%%%%%%%%
\newpage
\vspace{10mm}
\begin{center}
\begin{tabular}{c c c}
\multicolumn{3}{c}{TABLE 1. (Continued)}\\
  \hline
  \hline
Lie Algebra & Non Vanishing & Automorphism group \\&Structure Constants&\\
\hline
\hline
%%%%%%%%%%%%%%%%%%%%%%%%%%%%%%%%%%%%
\vspace{-3.5mm}
$g_{6,68}$\\
\vspace{-7.5mm} &$f^{1}_{35}=1,f^{2}_{45}=1,f^{1}_{16}=1$ &
$\left(\begin{array}{cccccc}
a_{3} & a_{4} & 0 & 0 & 0 & 0 \\
0 & a_{3} & 0 & 0 & 0 & 0 \\
a_{1} & a_{2} & a_{3} & a_{4} & 0 & 0\\
ca_{4} & a_{1} & 0 & a_{3} & 0 & 0 \\
a_{8}-a_{5} & a_{5} & 0 & 0 & 1 & 0 \\
a_{6} & a_{7} & a_{8}-a_{5} & a_{8}& 0 & 1\\
\end{array}
\right)$ \\
& $,f^{2}_{16}=1,f^{2}_{26}=1,f^{3}_{36}=1$\\
& $,f^{4}_{36}=1,f^{1}_{46}=c,f^{4}_{46}=1$\\
%%%%%%%%%%%%%%%%%%%%%%%%%%%%%%%%%%%%
\vspace{-3.5mm}
$g_{6,69}$\\
\end{tabular}
\end{center}
%%%%%%%%%%%%%%%%%%%%%%%%%%%%%%%%%%%%%%%%%%%%%%%%%%%%%%%%%%%%%%%%%%%%%%%%%%%%%%%%%%%%%%%%%%%%%%%%%%%%%%%%%
\newpage
\vspace{10mm}
\begin{center}
\begin{tabular}{c c c}
\multicolumn{3}{c}{TABLE 1. (Continued)}\\
  \hline
  \hline
Lie Algebra & Non Vanishing & Automorphism group \\&Structure Constants&\\
\hline \hline
%%%%%%%%%%%%%%%%%%%%%%%%%%%%%%%%%%%
\vspace{-3.5mm}
$g_{6,73}$\\
\vspace{-3.5mm} $\epsilon =1$
&$f^{1}_{25}=1,f^{2}_{35}=1,f^{3}_{45}=1,f^{1}_{16}=1,f^{2}_{26}=1$
& $\left(\begin{array}{cccccc}
a_{4} & 0 & 0 & 0 & 0 & 0 \\
a_{3} & a_{4} & 0 & 0 & 0 & 0 \\
a_{2} & a_{3} & a_{4} & 0 & 0 & 0\\
a_{1} & a_{2} & a_{3} & a_{4} & 0 & 0 \\
a_{6}-a_{8} & a_{7} & a_{8} & 0 & 1 & 0 \\
a_{5} & a_{6} & a_{7} & a_{8} & 0 & 1 \\
\end{array}
\right)$ \\
& $,f^{1}_{36}=1,f^{3}_{36}=1,f^{2}_{46}=1,f^{4}_{46}=1$\\
%%%%%%%%%%%%%%%%%%%%%%%%%%%%%%%%%%%%%%%%%%%%%%%%%%%%%%%%%%%
\vspace{-3.5mm}
$g_{6,73}$\\
\vspace{-3.5mm} $\epsilon =-1$
&$f^{1}_{25}=1,f^{2}_{35}=1,f^{3}_{45}=1,f^{1}_{16}=1,f^{2}_{26}=1$
& $\left(\begin{array}{cccccc}
a_{4} & 0 & 0 & 0 & 0 & 0 \\
a_{3} & a_{4} & 0 & 0 & 0 & 0 \\
a_{2} & a_{3} & a_{4} & 0 & 0 & 0\\
a_{1} & a_{2} & a_{3} & a_{4} & 0 & 0 \\
a_{6}+a_{8} & a_{7} & a_{8} & 0 & 1 & 0 \\
a_{5} & a_{6} & a_{7} & a_{8} & 0 & 1 \\
\end{array}
\right)$ \\
\end{tabular}
\end{center}
%%%%%%%%%%%%%%%%%%%%%%%%%%%%%%%%%%%%%%%%%%%%%%%%%%%%%%%%%%%%%%%%%%%%%%%%%%%%%%%%%%%%%%%%%%%%%%%%%%
\newpage
\vspace{10mm}
\begin{center}
\begin{tabular}{c c c}
\multicolumn{3}{c}{TABLE 1. (Continued)}\\
  \hline
  \hline
Lie Algebra & Non Vanishing & Automorphism group \\&Structure Constants&\\
\hline
 \hline
%%%%%%%%%%%%%%%%%%%%%%%%%%%%%%%%%%%
\vspace{-3.5mm}
$g_{6,78}$\\
\vspace{-3.5mm}
&$f^{3}_{24}=1,f^{1}_{25}=1,f^{2}_{45}=1,f^{1}_{16}=-1$ &
$\left(\begin{array}{cccccc}
a_{3} & 0 & 0 & 0 & 0 & 0 \\
a_{7} & 1 & -a_{6} & 0 & 0 & 0\\
0 & 0  & \frac{1}{a_{3}} & 0 & 0& 0 \\
\frac{a_{7}^{2}}{2a_{3}} & \frac{a_{7}}{a_{3}}& a_{1} & \frac{1}{a_{3}} & 0 & 0 \\
a_{2} & a_{3}a_{6} & -\frac{a_{3}a_{6}^{2}}{2} & 0 & a_{3} & 0 \\
a_{4} & a_{6}a_{7} & a_{5} & a_{6}& a_{7} & 1\\
\end{array}
\right)$ \\
& $,f^{3}_{36}=1,f^{3}_{46}=1,f^{4}_{46}=1,f^{5}_{56}=-1$\\
%%%%%%%%%%%%%%%%%%%%%%%%%%%%%%%%%%%%
\vspace{-3.5mm}
$g_{6,79}$\\
\vspace{-7.5mm}
&$f^{3}_{24}=1,f^{1}_{25}=1,f^{2}_{45}=1,f^{1}_{16}=3$ &
$\left(\begin{array}{cccccc}
a_{1}^{3} & 0 & 0 & 0 & 0 & 0 \\
(a_{6}-a_{7})a_{1}^{2} & a_{1}^{2} & -a_{6}a_{1}^{2} & 0 & 0 & 0 \\
a_{2}a_{1}^{2} & 0 & a_{1}^{3} & 0 & 0 & 0\\
\frac{2a_{2}a_{4}-2a_{2}a_{6}a_{7}+2a_{1}a_{7}^{2}-4a_{1}a_{6}a_{7}+a_{2}a_{6}^{2}+2a_{1}a_{6}^{2}}{4} & a_{1}a_{6}+a_{2}a_{6}-a_{1}a_{7}& \frac{2a_{1}a_{4}-2a_{2}a_{6}^{2}+2a_{1}a_{6}a_{7}-3a_{1}a_{6}^{2}}{4}  & a_{1} & a_{2} & 0 \\
\frac{2a_{1}a_{4}+a_{1}a_{6}^{2}-2a_{1}a_{6}a_{7}}{4}  & a_{1}a_{6} & -\frac{a_{1}a_{6}^{2}}{2} & 0 & a_{1} & 0 \\
a_{3} & a_{4} & a_{5} & a_{6}& a_{7} & 1\\
\end{array}
\right)$ \\
\end{tabular}
\end{center}
%%%%%%%%%%%%%%%%%%%%%%%%%%%%%%%%%%%%%%%%%%%%%%%%%%%%%%%%%%%%%%%%%%%%%%%%%%%%%%%%%%%%%%%%%%%%%%%%%%
\newpage
\vspace{10mm}
\begin{center}
\begin{tabular}{c c c}
\multicolumn{3}{c}{TABLE 1. (Continued)}\\
  \hline
  \hline
Lie Algebra & Non Vanishing & Automorphism group \\&Structure Constants&\\
\hline \hline
%%%%%%%%%%%%%%%%%%%%%%%%%%%%%%%%%%%
\vspace{-3.5mm}
$g_{6,82}$\\
\vspace{-3.5mm} $\alpha=0$
&$f^{1}_{24}=1,f^{1}_{35}=1,f^{2}_{26}=\lambda$ &
$\left(\begin{array}{cccccc}
a_{3}a_{7} & 0 & 0 & 0 & 0 & 0 \\
a_{1} & \frac{a_{3}a_{7}}{a_{5}} & 0 & 0 & 0 & 0 \\
a_{2} & 0 & a_{3} & 0 & 0 & 0\\
a_{4} & 0 & 0 & a_{5} & 0 & 0 \\
a_{6} & 0 & 0 & 0 & a_{7} & 0 \\
a_{8} & \frac{a_{4}\lambda}{a_{5}} & \frac{a_{6}\lambda_{1}}{a_{7}} & \frac{a_{1}a_{5}\lambda}{a_{3}a_{7}}& \frac{a_{2}\lambda_{1}}{a_{3}} & 1\\
\end{array}
\right)$ \\
& $,f^{3}_{36}=\lambda_{1},f^{4}_{46}=-\lambda,f^{5}_{56}=-\lambda_{1}$\\
%%%%%%%%%%%%%%%%%%%%%%%%%%%%%%%%%%%
\vspace{-3.5mm}
$g_{6,83}$\\
\vspace{-10.5mm} $\alpha\neq0$
&$f^{1}_{24}=1,f^{1}_{35}=1,f^{1}_{16}=\alpha$ &
$\left(\begin{array}{cccccc}
a_{2}a_{5} & 0 & 0 & 0 & 0 & 0 \\
a_{1} & a_{2} & -\frac{a_{2}a_{4}}{a_{5}} & 0 & 0 & 0 \\
\frac{2a_{2}a_{8}}{2\lambda-\alpha} & 0 & a_{2} & 0 & 0 & 0 \\
\frac{2a_{5}a_{7}}{2\lambda+\alpha} & 0 & 0 & a_{5} & 0 & 0\\
a_{3} & 0 & 0 & a_{4} & a_{5} & 0 \\
a_{6} & a_{7} &\frac{-2\alpha a_{4}a_{7}+4\alpha\lambda a_{3}+\alpha^{2}a_{3}-4\lambda a_{4}a_{7}+4a_{3}\lambda^{2}+4a_{5}a_{7}}{2a_{5}(\alpha+2\lambda)}&
\frac{-2\alpha a_{2}a_{4}a_{8}-4\alpha\lambda a_{1}a_{5}+4\lambda^{2}a_{1}a_{5}+\alpha^{2} a_{1}a_{5}+4\lambda a_{2}a_{4}a_{8}+4a_{2}a_{5}a_{8}}{2a_{2}a_{5}(2\lambda-\alpha)}& a_{8} & 1\\
\end{array}
\right)$ \\
\end{tabular}
\end{center}
%%%%%%%%%%%%%%%%%%%%%%%%%%%%%%%%%%%%%%%%%%%%%%%%%%%%%%%%%%%%%%%%%%%%%%%%%%%%%%%%%%%%%%%%%%%%%%%%%%
\newpage
\vspace{10mm}
\begin{center}
\begin{tabular}{c c c}
\multicolumn{3}{c}{TABLE 1. (Continued)}\\
  \hline
  \hline
Lie Algebra & Non Vanishing & Automorphism group \\&Structure Constants&\\
\hline \hline
%%%%%%%%%%%%%%%%%%%%%%%%%%%%%%%%%%%
\vspace{-3.5mm}
$g_{6,86}$\\
\vspace{-3.5mm}
&$f^{1}_{24}=1,f^{1}_{35}=1,f^{1}_{16}=2,f^{2}_{26}=1$ &
$\left(\begin{array}{cccccc}
a_{1}a_{5} & 0 & 0 & 0 & 0 & 0 \\
\frac{a_{2}a_{5}a_{7}-a_{1}a_{5}a_{9}-a_{1}a_{5}a_{10}+a_{1}a_{4}a_{10}}{a_{5}}& a_{1} & -\frac{a_{1}a_{4}}{a_{5}} & a_{2} & 0 & 0 \\
-a_{1}a_{10} & 0 & a_{1} & 0 & 0 & 0 \\
a_{5}a_{7} & 0 & 0 & a_{5} & 0 & 0 \\
a_{4}a_{7}-a_{5}a_{7}+a_{5}a_{8}-a_{3}a_{10} & 0 & a_{3} & a_{4} & a_{5} & 0\\
a_{6} &  a_{7} &  a_{8} & a_{9} & a_{10} & 1\\
\end{array}
\right)$ \\
& $,f^{3}_{26}=1,f^{3}_{36}=1,f^{4}_{46}=1,f^{4}_{56}=-1,f^{5}_{56}=1$\\
%%%%%%%%%%%%%%%%%%%%%%%%%%%%%%%%%%%
\vspace{-3.5mm}
$g_{6,87}$\\
\vspace{-3.5mm}
&$f^{1}_{24}=1,f^{1}_{35}=1,f^{1}_{16}=2,f^{2}_{26}=1,f^{5}_{26}=1$
& $\left(\begin{array}{cccccc}
a_{1}^{2} & 0 & 0 & 0 & 0 & 0 \\
\frac{2a_{1}a_{2}a_{5}+2a_{1}a_{3}a_{6}-2a_{1}^{2}a_{7}-a_{3}^{2}a_{8}+2a_{1}^{2}a_{5}+a_{3}^{2}a_{5}-2a_{1}a_{3}a_{8}+2a_{1}^{2}a_{6}-2a_{1}^{2}a_{8}+2a_{1}a_{3}a_{5}}{2a_{1}}& a_{1} & \frac{a_{3}^{2}}{2a_{1}} & a_{2} & a_{3} & 0 \\
a_{1}a_{5}-a_{1}a_{8}+a_{3}a_{5} & 0 & a_{1} & a_{3}& 0 & 0 \\
a_{1}a_{5} & 0 & 0 & a_{1} & 0 & 0 \\
\frac{2a_{1}^{2}a_{5}+a_{3}^{2}a_{5}-2a_{1}a_{3}a_{8}+2a_{1}^{2}a_{6}-2a_{1}^{2}a_{8}+2a_{1}a_{3}a_{5}}{2a_{1}}& 0  & a_{3} & \frac{a_{3}^{2}}{2a_{1}}& a_{1} & 0 \\
a_{4} &  a_{5} &  a_{6} & a_{7} & a_{8} & 1\\
\end{array}
\right)$ \\
\end{tabular}
\end{center}
%%%%%%%%%%%%%%%%%%%%%%%%%%%%%%%%%%%%%%%%%%%%%%%%%%%%%%%%%%%%%%%%%%%%%%%%%%%%%%%%%%%%%%%%%%%%%%%%%%%%%
\newpage
\vspace{10mm}
\begin{center}
\begin{tabular}{c c c}
\multicolumn{3}{c}{TABLE 1. (Continued)}\\
  \hline
  \hline
Lie Algebra & Non Vanishing & Automorphism group \\&Structure Constants&\\
\hline \hline
%%%%%%%%%%%%%%%%%%%%%%%%%%%%%%%%%%%
\vspace{-3.5mm}
$g_{6,89}$\\
\vspace{-3.5mm} $\alpha =0$
&$f^{1}_{24}=1,f^{1}_{35}=1,f^{2}_{26}=s,f^{5}_{36}=\nu_{0}$ &
$\left(\begin{array}{cccccc}
 a_{3}^{2}+a_{4}^{2} & 0 & 0 & 0 & 0 & 0 \\
\frac{a_{7}(a_{3}^{2}+a_{4}^{2})}{a_{2}s}& \frac{a_{3}^{2}+a_{4}^{2}}{a_{2}}& 0 & 0 & 0 & 0 \\
\frac{a_{4}a_{6}-a_{3}a_{8}}{\nu_{0}}& 0 & a_{4}  & 0 & -a_{3} & 0 \\
a_{1}& 0 & 0 & a_{2}& 0 & 0  \\
\frac{a_{3}a_{6}+a_{4}a_{8}}{\nu_{0}}& 0 & a_{3}  & 0 & a_{4} & 0 \\
a_{5} &  \frac{a_{1}s}{a_{2}} &  a_{6} & a_{7} & a_{8} & 1\\
\end{array}
\right)$ \\
& $,f^{4}_{46}=-s,f^{3}_{56}=-\nu_{0}$\\
%%%%%%%%%%%%%%%%%%%%%%%%%%%%%%%%%%%
\vspace{-3.5mm}
$g_{6,90}$\\
\vspace{-7.5mm}
 $\alpha\neq 0$
&$f^{1}_{24}=1,f^{1}_{35}=1,f^{1}_{16}=\alpha,f^{2}_{26}=\frac{\alpha}{2}$
& $\left(\begin{array}{cccccc}
a_{1}^{2}-a_{2}^{2} & 0 & 0 & 0 & 0 & 0 \\
\frac{2(2a_{1}a_{5}-2a_{2}a_{7}+\alpha a_{2}a_{5}-\alpha a_{1}a_{7})}{\alpha^{2}-4} & a_{1} & 0 & a_{2} & 0 & 0 \\
-\frac{2(r\alpha a_{6}+\alpha a_{3}a_{8}-2\nu_{0}a_{3}a_{6}+2r\nu_{0}a_{8})}{\alpha^{2}+4\nu_{0}^{2}} & 0 & a_{3} & 0 & -r & 0\\
\frac{2(2a_{2}a_{5}-2a_{1}a_{7}+\alpha a_{1}a_{5}-\alpha a_{2}a_{7})}{\alpha^{2}-4} & a_{2} & 0 & a_{1} & 0 & 0 \\
-\frac{2(r\alpha a_{8}-\alpha a_{3}a_{6}-2\nu_{0}a_{3}a_{8}-2r\nu_{0}a_{6})}{\alpha^{2}+4\nu_{0}^{2}} & 0 & r & 0 & a_{3} & 0\\
a_{4} & a_{5} & a_{6} & a_{7}& a_{8} & 1\\
\end{array}
\right)$ \\
\end{tabular}
\end{center}
%%%%%%%%%%%%%%%%%%%%%%%%%%%%%%%%%%%%%%%%%%%%%%%%%%%%%%%%%%%%%%%%%%%%%%%%%%%%%%%%%%%%%%%%%%%%%%%%%%%%%
\newpage
\vspace{10mm}
\begin{center}
\begin{tabular}{c c c}
\multicolumn{3}{c}{TABLE 1. (Continued)}\\
  \hline
  \hline
Lie Algebra & Non Vanishing & Automorphism group \\&Structure Constants&\\
\hline \hline
%%%%%%%%%%%%%%%%%%%%%%%%%%%%%%%%%%%
\vspace{-3.5mm}
$g^{*}_{6,92}$\\
\vspace{-8.5mm}$p\neq0$ &$f^{5}_{24}=1,f^{5}_{13}=1,f^{1}_{16}=p$
& $\left(\begin{array}{cccccc}
r & 0 & a_{1} & 0 & \frac{p a_{1}a_{4}+r a_{4}+a_{1}a_{6}-p r a_{6}}{1+p^{2}} & 0 \\
0 & a_{2} & 0 & a_{3} & \frac{a_{3}a_{7}+p a_{3}a_{5}-p a_{2}a_{7}+ a_{2}a_{5}}{1+p^{2}} & 0 \\
-a_{1} & 0 & r & 0 & \frac{-a_{1}a_{4}+r a_{6}+p r a_{4}+p a_{1}a_{6}}{1+p^{2}} & 0 \\
0 & -a_{3} & 0 & a_{2} & \frac{p a_{2}a_{5}+p a_{3}a_{7}+a_{2}a_{7}-a_{3}a_{5}}{1+p^{2}} & 0 \\
0 & 0 & 0 & 0 & a_{2}^{2}+a_{3}^{2} & 0 \\
a_{4} & a_{5} & a_{6} & a_{7}& a_{8} & 1\\
\end{array}
\right)$\\
 &$,f^{3}_{16}=1,f^{2}_{26}=p,f^{4}_{26}=1$\\
& $,f^{1}_{36}=-1,f^{3}_{36}=p,f^{2}_{46}=-1$\\
 & $,f^{4}_{46}=p,f^{5}_{56}=2p$& $r=Rootof(z^{2}+a_{1}^{2}-a_{2}^{2}-a_{3}^{2})$\\
\end{tabular}
\end{center}
%%%%%%%%%%%%%%%%%%%%%%%%%%%%%%%%%%%%%%%%%%%%%%%%%%%%%%%%%%%%%%%%%%%%%%%%%%%%%%%%%%%%%%%%%%%%%%%%%%%%%
\newpage
\vspace{10mm}
\begin{center}
\begin{tabular}{c c c}
\multicolumn{3}{c}{TABLE 1. (Continued)}\\
  \hline
  \hline
Lie Algebra & Non Vanishing & Automorphism group \\&Structure Constants&\\
\hline \hline
%%%%%%%%%%%%%%%%%%%%%%%%%%%%%%%%%%%
\vspace{-3.5mm}
$g_{6,95}$\\
\vspace{-3.5mm}
 &$f^{1}_{34}=1,f^{1}_{25}=1,f^{2}_{35}=1,f^{1}_{16}=2$ &
$\left(\begin{array}{cccccc}
a_{3}^{2} & 0 & 0 & 0 & 0 & 0 \\
-a_{3}a_{8}& a_{3} & 0 & 0 & 0 & 0 \\
\frac{a_{8}^{2}-a_{7}}{2}& -a_{8} & 1 & 0 & 0 & 0 \\
a_{1}& 0 & 0 & a_{3}^{2} & 0 & 0 \\
a_{3}a_{5}-a_{2}a_{8}& a_{2} & 0 & 0 & a_{3} & 0 \\
a_{4} &  a_{5} &  0 & a_{7} & a_{8} & 1\\
\end{array}
\right)$ \\
& $,f^{2}_{26}=1,f^{1}_{46}=1,f^{4}_{46}=2,f^{5}_{56}=1$\\
%%%%%%%%%%%%%%%%%%%%%%%%%%%%%%%%%%%
\vspace{-3.5mm}
$g_{6,96}$\\
\vspace{-3.5mm}
&$f^{1}_{34}=1,f^{1}_{25}=1,f^{2}_{35}=1,f^{1}_{16}=3,f^{2}_{26}=2$
& $\left(\begin{array}{cccccc}
a_{2}^{3} & 0 & 0 & 0 & 0 & 0 \\
-a_{2}^{2}a_{7}& a_{2}^{2} & 0 & 0 & 0 & 0 \\
\frac{a_{7}^{2}a_{2}-a_{6}a_{2}}{2}& -a_{2}a_{7} & a_{2} & 0 & 0 & 0 \\
a_{2}^{2}a_{5}-a_{1}a_{2}a_{7}-a_{2}^{2}a_{7}& a_{1}a_{2} & 0 & a_{2}^{2} & 0 & 0 \\
\frac{2a_{2}a_{4}-2a_{1}a_{6}-a_{2}a_{6}-2a_{2}a_{5}a_{7}+2a_{1}a_{7}^{2}+3a_{2}a_{7}^{2}}{4}& a_{2}a_{5}-a_{1}a_{7}-a_{2}a_{7} & a_{1} & 0 & a_{2} & 0 \\
a_{3} & a_{4} &  a_{5} & a_{6} & a_{7} & 1\\
\end{array}
\right)$ \\
\end{tabular}
\end{center}
%%%%%%%%%%%%%%%%%%%%%%%%%%%%%%%%%%%%%%%%%%%%%%%%%%%%%%%%%%%%%%%%%%%%%%%%%%%%%%%%%%%%%%%%%%%%%%%%%%%%%%%%%%%%%%%%%%
\newpage
\vspace{10mm}
\begin{center}
\begin{tabular}{c c c}
\multicolumn{3}{c}{TABLE 2: The automorphism groups of six dimensional solvable real Lie algebras with nilradical 4}\\
  \hline
  \hline
Lie Algebra & Non Vanishing & Automorphism group \\&Structure Constants&\\
\hline \hline \vspace{-4mm}
$N^{\alpha\beta\gamma\delta}_{6,1}$\\
\vspace{-4mm} $\alpha\beta\neq0$
&$f^{3}_{13}=\alpha,f^{4}_{14}=\gamma,f^{6}_{16}=1,f^{3}_{23}=\beta$
& $\left(\begin{array}{cccccc}
1 & 0 & \frac{a_{2}\alpha}{\beta} & \frac{a_{3}\gamma}{\delta} & 0 & a_{1} \\
0 & 1 & a_{2} & a_{3} & a_{4} & 0 \\
0 & 0 & a_{5} & 0 & 0 & 0 \\
0 & 0 & 0 & a_{6}& 0 & 0 \\
0 & 0 & 0& 0 & a_{7} & 0\\
0 & 0 & 0 & 0 & 0 & a_{8}\\
\end{array}
\right)$ \\
$\gamma^{2}+\delta^{2}\neq0$& $,f^{4}_{24}=\delta , f^{5}_{25}=1 $\\
%%%%%%%%%%%%%%%%%%%%%%%%%%%%%%%%%
\vspace{-4mm}
$N^{\alpha\beta\gamma}_{6,2}$\\
\vspace{-4mm} $\alpha^{2}+\beta^{2}\neq0$
&$f^{3}_{13}=\alpha,f^{4}_{14}=1,f^{6}_{15}=1,f^{3}_{23}=\beta$ &
$\left(\begin{array}{cccccc}
1 & 0 & \frac{a_{3}\alpha}{\beta} & a_{1} & 0 & a_{2} \\
0 & 1 & a_{3} & \gamma a_{1} & a_{2} & a_{4} \\
0 & 0 & a_{5} & 0 & 0 & 0 \\
0 & 0 & 0 & a_{6}& 0 & 0 \\
0 & 0 & 0& 0 & a_{7} & a_{8}\\
0 & 0 & 0 & 0 & 0 & a_{7}\\
\end{array}
\right)$ \\
\end{tabular}
\end{center}
%%%%%%%%%%%%%%%%%%%%%%%%%%%%%%%%%%%
\newpage
\vspace{10mm}
\begin{center}
\begin{tabular}{c c c}
\multicolumn{3}{c}{TABLE 2. (Continued)}\\
  \hline
  \hline
Lie Algebra & Non Vanishing & Automorphism group \\&Structure Constants&\\
\hline \hline
%%%%%%%%%%%%%%%%%%%%%%%%%%%%%%%%%%%%%
\vspace{-4mm}
$N^{\alpha\beta\gamma}_{6,7}$\\
\vspace{-4mm} $\alpha^{2}+\beta^{2}\neq0$
&$f^{3}_{13}=\alpha,f^{4}_{14}=\alpha,f^{5}_{15}=1,f^{6}_{15}=1,f^{6}_{16}=1$
& $\left(\begin{array}{cccccc}
1 & 0 & \frac{\alpha(\gamma a_{3}+a_{4})}{1+\gamma^{2}} & \frac{\alpha(\gamma a_{4}-a_{3})}{1+\gamma^{2}} & a_{1} & a_{2} \\
0 & 1 & a_{3} & a_{4} & 0 & \beta a_{1} \\
0 & 0 & a_{5} & a_{6} & 0 & 0 \\
0 & 0 & -a_{6} & a_{5}& 0 & 0 \\
0 & 0 & 0& 0 & a_{7} & a_{8}\\
0 & 0 & 0 & 0 & 0 & a_{7}\\
\end{array}
\right)$ \\
& $,f^{3}_{23}=\gamma,f^{4}_{23}=1,f^{3}_{24}=-1,f^{4}_{24}=\gamma,f^{6}_{25}=\beta$\\
%%%%%%%%%%%%%%%%%%%%%%%%%%%%%%%%%%%
\vspace{-4mm}
$N_{6,8}$\\
\vspace{-4mm}
&$f^{3}_{13}=1,f^{6}_{14}=1,f^{4}_{24}=1,f^{5}_{25}=1$ &
$\left(\begin{array}{cccccc}
1 & 0 & a_{1} & 0 & 0 & a_{2} \\
0 & 1 & 0 & a_{2} & a_{3} & a_{4} \\
0 & 0 & a_{5} & 0 & 0 & 0 \\
0 & 0 & 0 & a_{6}& 0 & a_{7}\\
0 & 0 & 0& 0 & a_{6} & a_{8}\\
0 & 0 & 0 & 0 & 0 & a_{6}\\
\end{array}
\right)$ \\
\end{tabular}
\end{center}
%%%%%%%%%%%%%%%%%%%%%%%%%%%%%%%%%%
\newpage
\vspace{10mm}
\begin{center}
\begin{tabular}{c c c}
\multicolumn{3}{c}{TABLE 2. (Continued) }\\
  \hline
  \hline
Lie Algebra & Non Vanishing & Automorphism group \\&Structure Constants&\\
\hline \hline
%%%%%%%%%%%%%%%%%%%%%%%%%%%%%%%%%%%
\vspace{-4mm}
$N^{\alpha\beta\gamma\delta}_{6,13}$\\
\vspace{-4mm}
 $\alpha^{2}+\beta^{2}\neq0$
&$f^{3}_{13}=\alpha,f^{4}_{14}=\gamma,f^{6}_{15}=1,f^{5}_{16}=-1$
& $\left(\begin{array}{cccccc}
1 & 0 & a_{1} & \frac{a_{3}\gamma}{\delta} & -a_{4} & a_{2} \\
0 & 1 & \frac{a_{1}\beta}{\alpha} & a_{3} & a_{2} & a_{4} \\
0 & 0 & a_{5} & 0 & 0 & 0 \\
0 & 0 & 0 & a_{6}& 0 & 0 \\
0 & 0 & 0& 0 & a_{7} & a_{8}\\
0 & 0 & 0 & 0 & -a_{8} & a_{7}\\
\end{array}
\right)$ \\
$\gamma^{2}+\delta^{2}\neq0$& $,f^{3}_{23}=\beta,f^{4}_{24}=\delta , f^{5}_{25}=1,f^{6}_{26}=1 $\\
%%%%%%%%%%%%%%%%%%%%%%%%%%%%%%%%%%%
 \vspace{-4mm}
$N^{\alpha\beta\gamma}_{6,14}$\\
\vspace{-3.5mm}
 $\alpha\beta\neq0$
&$f^{3}_{13}=\alpha,f^{5}_{15}=\gamma,f^{6}_{15}=1,f^{5}_{16}=-1$
& $\left(\begin{array}{cccccc}
1 & 0 & \frac{a_{3}\alpha}{\beta} & 0 & a_{1} & a_{2} \\
0 & 1 & a_{3} & a_{4}& 0  & 0 \\
0 & 0 & a_{5} & 0 & 0 & 0 \\
0 & 0 & 0 & a_{6}& 0 & 0 \\
0 & 0 & 0& 0 & a_{7} & a_{8}\\
0 & 0 & 0 & 0 & -a_{8} & a_{7}\\
\end{array}
\right)$ \\
\end{tabular}
\end{center}
%%%%%%%%%%%%%%%%%%%%%%%%%%%%%%%%%%
\newpage
\vspace{10mm}
\begin{center}
\begin{tabular}{c c c}
\multicolumn{3}{c}{TABLE 2. (Continued)}\\
  \hline
  \hline
Lie Algebra & Non Vanishing & Automorphism group \\&Structure Constants&\\
\hline \hline
%%%%%%%%%%%%%%%%%%%%%%%%%%%%%%%%%%%%%%%%%%%
\vspace{-4.8mm}
$N_{6,19}$\\
\vspace{-4.8mm}
&$f^{4}_{13}=1,f^{5}_{13}=1,f^{3}_{14}=-1,f^{6}_{14}=1,f^{6}_{15}=1$
& $\left(\begin{array}{cccccc}
1 & 0 & -a_{3} & a_{1}+a_{4} & a_{1} & a_{2} \\
0 & 1 & a_{1}+a_{4} & a_{3} & a_{2}-a_{3} & a_{4} \\
0 & 0 & a_{5} & -a_{8} & a_{7} & -a_{6} \\
0 & 0 & a_{8} & a_{5} & a_{6} & a_{7} \\
0 & 0 & 0 & 0 & a_{5} & -a_{8} \\
0 & 0 & 0 & 0 & a_{8} & a_{5} \\
\end{array}
\right)$ \\
& $,f^{5}_{16}=-1,f^{3}_{23}=1,f^{4}_{24}=1,f^{5}_{25}=1,f^{6}_{26}=1$\\
%%%%%%%%%%%%%%%%%%%%%%%%%%%%%%%%%%%
\vspace{-5mm}
$N^{\alpha\beta}_{6,20}$\\
 $\alpha^{2}+\beta^{2}\neq0$
&$f^{4}_{14}=\alpha,f^{6}_{16}=1,f^{4}_{24}=\beta,f^{5}_{25}=1,f^{3}_{12}=1$
& $\left(\begin{array}{cccccc}
1 & 0 & a_{1}& \frac{\alpha a_{4}}{\beta}  & 0 & a_{2} \\
0 & 1 & a_{3} & a_{4} &  a_{5} & 0 \\
0 & 0 & 1 & 0 & 0 & 0 \\
0 & 0 & 0 & a_{6} & 0 & 0 \\
0 & 0 & 0 &  0 & a_{7} & 0 \\
0 & 0 & 0 & 0 & 0 & a_{8}\\
\end{array}
\right)$ \\
\end{tabular}
\end{center}
%%%%%%%%%%%%%%%%%%%%%%%%%%%%%%%%%%%
\newpage
\vspace{10mm}
\begin{center}
\begin{tabular}{c c c}
\multicolumn{3}{c}{TABLE 2. (Continued)}\\
  \hline
  \hline
Lie Algebra & Non Vanishing & Automorphism group \\&Structure Constants&\\
\hline \hline
%%%%%%%%%%%%%%%%%%%%%%%%%%%%%%%%%%%%
\vspace{-4mm}
$N_{6,24}$\\
\vspace{-3.5mm} &$f^{5}_{15}=1,f^{6}_{15}=1,f^{6}_{16}=1$ &
$\left(\begin{array}{cccccc}
1 & 0 & a_{1} &  0 &  a_{2} &  a_{3} \\
0 & 1 &  a_{4} & a_{5} &  0 &  0 \\
0 & 0 & 1 &  0 & 0 & 0 \\
0 & 0 & 0 &  a_{6} & 0 & 0 \\
0 & 0 & 0& 0 & a_{7} & a_{8} \\
0 & 0 & 0& 0& 0 & a_{7}\\
\end{array}
\right)$ \\
& $,f^{4}_{24}=1,f^{3}_{12}=1$\\
%%%%%%%%%%%%%%%%%%%%%%%%%%%%%%%%%%%%
\vspace{-4mm}
$N^{\alpha\beta}_{6,25}$\\
\vspace{-3.5mm} $\alpha^{2}+\beta^{2}\neq0$
&$f^{4}_{14}=\alpha,f^{6}_{15}=1,f^{5}_{16}=-1,f^{4}_{23}=\beta$ &
$\left(\begin{array}{cccccc}
1 & 0 & \frac{\alpha a_{3}}{\beta} &  a_{1} &  -a_{5} &  a_{4} \\
0 & 1 &  a_{2} & a_{3} &  a_{4} &  a_{5} \\
0 & 0 & 1 &  0 & 0 & 0 \\
0 & 0 & 0 &  1 & 0 & 0 \\
0 & 0 & 0& 0 & a_{7} & -a_{6} \\
0 & 0 & 0& 0& a_{6} & a_{7}\\
\end{array}
\right)$ \\
\end{tabular}
\end{center}
%%%%%%%%%%%%%%%%%%%%%%%%%%%%%%%%%%%
\newpage
\vspace{10mm}
\begin{center}
\begin{tabular}{c c c}
\multicolumn{3}{c}{TABLE 2. (Continued)}\\
  \hline
  \hline
Lie Algebra & Non Vanishing & Automorphism group \\&Structure Constants&\\
\hline \hline
%%%%%%%%%%%%%%%%%%%%%%%%%%%%%%%%%%%%
\vspace{-4mm}
$N^{\alpha\beta}_{6,29}$\\
\vspace{-4mm} $\alpha^{2}+\beta^{2}\neq0$
&$f^{3}_{45}=1,f^{3}_{13}=1 ,f^{4}_{14}=1,f^{6}_{16}=\alpha$ &
$\left(\begin{array}{cccccc}
1 & 0 & a_{2}+a_{1}a_{3} & a_{1} & 0 & \frac{\alpha a_{4}}{\beta} \\
0 & 1 & a_{2} & 0 & a_{3} & a_{4} \\
0 & 0 & a_{5}a_{6} & 0 & 0 & 0 \\
0 & 0 & a_{3}a_{5} & a_{5}& 0 & 0 \\
0 & 0 & -a_{1}a_{6}& 0 & a_{6} & 0\\
0 & 0 & 0 & 0 & 0 & a_{7}\\
\end{array}
\right)$ \\
& $,f^{3}_{23}=1,f^{5}_{25}=1,f^{6}_{26}=\beta$\\
%%%%%%%%%%%%%%%%%%%%%%%%%%%%%%%%%%%
\vspace{-4mm}
$N^{\alpha}_{6,30}$\\
\vspace{-4mm}
&$f^{3}_{45}=1,f^{3}_{13}=2,f^{4}_{14}=1,f^{5}_{15}=1$ &
$\left(\begin{array}{cccccc}
1 & 0 & a_{1} & a_{3} & a_{2} & \alpha a_{4} \\
0 & 1 & -\frac{a_{3}^{2}}{2} & 0 & a_{3} & a_{4} \\
0 & 0 & a_{5}^{2} & 0 & 0 & 0 \\
0 & 0 & a_{2}a_{5}-a_{3}a_{6} & a_{5}& a_{6} & 0 \\
0 & 0 & -a_{3}a_{5}& 0 & a_{5} & 0\\
0 & 0 & 0 & 0 & 0 & a_{7}\\
\end{array}
\right)$ \\
\end{tabular}
\end{center}
%%%%%%%%%%%%%%%%%%%%%%%%%%%%%%%%%%%
\newpage
\vspace{10mm}
\begin{center}
\begin{tabular}{c c c}
\multicolumn{3}{c}{TABLE 2. (Continued)}\\
  \hline
  \hline
Lie Algebra & Non Vanishing & Automorphism group \\&Structure Constants&\\
\hline \hline
%%%%%%%%%%%%%%%%%%%%%%%%%%%%%%%%%%%
\vspace{-4mm}
$N^{\alpha\beta}_{6,35}$\\
\vspace{-4mm} $\alpha^{2}+\beta^{2}\neq0$
&$f^{3}_{45}=1,f^{5}_{14}=1,f^{4}_{15}=-1,f^{6}_{16}=\alpha,f^{3}_{23}=2$
& $\left(\begin{array}{cccccc}
1 & 0 & -\frac{a_{2}^{2}+a_{3}^{2}}{2} & -a_{3} & a_{2} &\frac{ \alpha a_{4}}{\beta}  \\
0 & 1 & a_{1} & a_{2} & a_{3} & a_{4} \\
0 & 0 & a_{5}^{2}+a_{6}^{2} & 0 & 0 & 0 \\
0 & 0 & a_{3}a_{6}+a_{2}a_{5}& a_{6} & -a_{5} & 0\\
0 & 0 & a_{3}a_{5}-a_{2}a_{6}& a_{5} & a_{6} & 0\\
0 & 0 & 0 & 0 & 0 & a_{7}\\
\end{array}
\right)$ \\
& $,f^{4}_{24}=1 ,f^{5}_{25}=1,f^{6}_{26}=\beta$\\
%%%%%%%%%%%%%%%%%%%%%%%%%%%%%%%%%%%
\vspace{-3.5mm}
$N_{6,36}$\\
\vspace{-4mm}
&$f^{3}_{45}=1,f^{5}_{14}=1,f^{4}_{15}=-1,f^{3}_{23}=2$ &
$\left(\begin{array}{cccccc}
1 & 0 & -\frac{a_{2}^{2}+a_{3}^{2}}{2} & -a_{3} & a_{2} & 0  \\
0 & 1 & a_{1} & a_{2} & a_{3} & a_{4} \\
0 & 0 & a_{5}^{2}+a_{6}^{2} & 0 & 0 & 0 \\
0 & 0 & a_{3}a_{6}+a_{2}a_{5}& a_{6} & -a_{5} & 0\\
0 & 0 & a_{3}a_{5}-a_{2}a_{6}& a_{5} & a_{6} & 0\\
0 & 0 & a_{7} & 0 & 0 & a_{5}^{2}+a_{6}^{2}\\
\end{array}
\right)$ \\
\end{tabular}
\end{center}
%%%%%%%%%%%%%%%%%%%%%%%%%%%%%%%%%%
\newpage
\begin{center}
\begin{tabular}{c c c}
\multicolumn{3}{c}{TABLE 3: The automorphism groups of six dimensional nilpotent real Lie algebras   }\\
  \hline
  \hline
Lie Algebra & Non Vanishing & Automorphism group \\&Structure Constants&\\
\hline \hline
%%%%%%%%%%%%%%%%%%%%%%%%%%%%%%%%%%%%%%%%%%%%%%
\vspace{-3.5mm}
$A_{6,1}$\\
&$f^{3}_{12}=1,f^{4}_{13}=1,f^{6}_{15}=1$ &
$\left(\begin{array}{cccccc}
a_{1} & a_{2} & a_{3} & a_{4} & a_{5} & a_{6} \\
0 & a_{7} & a_{8} & a_{9} & a_{10} & a_{11} \\
0 & 0 & a_{1}a_{7} & a_{1}a_{8} & 0 & a_{1}a_{10} \\
0 & 0 & 0 & a_{1}^{2}a_{7}& 0 & 0 \\
0 & 0 & a_{12}& a_{13}& a_{14} & a_{15}\\
0 & 0 & 0& a_{1}a_{12} & 0 & a_{1}a_{14}\\
\end{array}
\right)$\\
%%%%%%%%%%%%%%%%%%%%%%%%%%%%%%
\vspace{-4mm}
$A_{6,2}$\\
&$f^{3}_{12}=1,f^{4}_{13}=1,f^{5}_{14}=1,f^{6}_{15}=1$ &
$\left(\begin{array}{cccccc}
a_{1} & a_{2} & a_{3} & a_{4} & a_{5} & a_{6} \\
0 & a_{7} & a_{8} & a_{9} & a_{10} & a_{11} \\
0 & 0 & a_{1}a_{7} & a_{1}a_{8} & a_{1}a_{9} & a_{1}a_{10} \\
0 & 0 & 0 & a_{1}^{2}a_{7}& a_{1}^{2}a_{8} & a_{1}^{2}a_{9} \\
0 & 0 & 0 & 0 & a_{1}^{3}a_{7} & a_{1}^{3}a_{8}\\
0 & 0 & 0& 0 & 0 & a_{1}^{4}a_{7}\\
\end{array}
\right)$  \\
%%%%%%%%%%%%%%%%%%%%%%%%%%%%%%%
\vspace{-3.5mm}
$A_{6,3}$\\
\end{tabular}
\end{center}
%%%%%%%%%%%%%%%%%%%%%%%%%%%%%%%%%%%%%%%%%%%%%%%%%%%%%%%%%%%%%%%%%%%%%%%%%%%%%%%%%%%%%%%%%%%%%%%%%%%%%%%%%%%%%%%%%%%%%%%%%%%%
\newpage
\vspace{10mm}
\begin{center}
\begin{tabular}{c c c}
\multicolumn{3}{c}{TABLE 3. (Continued) }\\
  \hline
  \hline
Lie Algebra & Non Vanishing & Automorphism group \\&Structure Constants&\\
\hline \hline
%%%%%%%%%%%%%%%%%%%%%%%%%%%%%%%%%%%%%%%%%%%%%%
\vspace{-3.5mm}
$A_{6,6}$\\
&$f^{6}_{12}=1,f^{4}_{13}=1,f^{5}_{14}=1,f^{5}_{23}=1$ &
$\left(\begin{array}{cccccc}
a_{1} & a_{2} & a_{3} & a_{4} & a_{5} & a_{6} \\
0 & a_{1}^{2} & 0 & a_{7} & a_{8} & a_{9} \\
0 & a_{10} & a_{11} & a_{12} & a_{13} & a_{14} \\
0 & 0 & 0 & a_{1}a_{11} & a_{2}a_{11}+a_{1}a_{12}-a_{3}a_{10} & a_{1}a_{10} \\
0 & 0 & 0 & 0 & a_{1}^{2}a_{11} & 0 \\
0 & 0 & 0 & 0 & a_{1}a_{7}-a_{1}^{2}a_{3} & a_{1}^{3}\\
\end{array}
\right)$  \\
%%%%%%%%%%%%%%%%%%%%%%%%%%%%%%%%
\vspace{-3.5mm}
$A_{6,7}$\\
&$f^{4}_{13}=1,f^{5}_{14}=1,f^{6}_{23}=1$ &
$\left(\begin{array}{cccccc}
a_{1} & a_{2} & 0 & a_{3} & a_{4} & a_{5} \\
0 & a_{6} & 0 & 0 & a_{7} & a_{8} \\
0 & a_{9} & a_{10} & a_{11} & a_{12} & a_{13} \\
0 & 0 & 0 & a_{1}a_{10}& a_{1}a_{11} & a_{2}a_{10} \\
0 & 0 & 0 & 0 & a_{1}^{2}a_{10} & 0\\
0 & 0 & 0 & 0 & 0 & a_{6}a_{10}\\
\end{array}
\right)$\\
%%%%%%%%%%%%%%%%%%%%%%%%%%%%%%
\vspace{-3.5mm}
$A_{6,8}$\\
\end{tabular}
\end{center}
%%%%%%%%%%%%%%%%%%%%%%%%%%%%%%%%%%%%%%%%%%%%%%%%%%%%%%%%%%%%%%%%%%%%%%%%%%%%%%%%%%%%%%%%%%%%%%%%%%%%%%%%%%%%%%%%%%%%%%%%%%%%
\newpage
\vspace{10mm}
\begin{center}
\begin{tabular}{c c c}
\multicolumn{3}{c}{TABLE 3. (Continued)}\\
\hline \hline
Lie Algebra & Non Vanishing & Automorphism group \\&Structure Constants&\\
\hline \hline
%%%%%%%%%%%%%%%%%%%%%%%%%%%%%%%%%%%%%%%%%%%%%%
\vspace{-3.5mm}
$A_{6,12}$\\
&$f^{4}_{13}=1,f^{6}_{14}=1,f^{6}_{25}=1$ &
$\left(\begin{array}{cccccc}
a_{1} & a_{2} & a_{3} & a_{4} & a_{5} & a_{6} \\
0 & a_{7} & 0 & \frac{a_{5}a_{7}-a_{2}a_{8}}{a_{1}} & a_{8} &  a_{9} \\
0  & 0 & \frac{a_{7}a_{13}-a_{8}a_{12}}{a_{1}^{2}}& a_{10} & 0 &  a_{11} \\
0 & 0 & 0 & \frac{a_{7}a_{13}-a_{8}a_{12}}{a_{1}} & 0  & a_{1}a_{10} \\
0 & a_{12} & 0 & \frac{a_{5}a_{12}-a_{2}a_{13}}{a_{1}}& a_{13} & a_{14} \\
0 & 0 & 0 & 0 & 0 & a_{7}a_{13}-a_{8}a_{12}\\
\end{array}
\right)$  \\
%%%%%%%%%%%%%%%%%%%%%%%%%%%%%%%%
\vspace{-3.5mm}
$A_{6,13}$\\
&$f^{5}_{12}=1,f^{4}_{13}=1,f^{6}_{14}=1,f^{6}_{25}=1$ &
$\left(\begin{array}{cccccc}
\frac{a_{6}^{2}}{a_{10}} & a_{1} & a_{2} & a_{3} & a_{4} & a_{5} \\
0 & a_{6} & -\frac{a_{1}a_{10}}{a_{6}} & a_{7} & a_{8} & a_{9} \\
0 & 0 & a_{10} & a_{11} & 0 & a_{12}  \\
0 & 0 & 0 & a_{6}^{2} & 0 & \frac{a_{11}a_{6}^{2}}{a_{10}} \\
0 & 0 & 0 & -a_{1}a_{6} & \frac{a_{6}^{3}}{a_{10}} & \frac{a_{6}^{2}a_{7}+a_{1}a_{8}a_{10}-a_{4}a_{6}a_{10}}{a_{10}}\\
0 & 0 & 0 & 0 & 0 &  \frac{a_{6}^{4}}{a_{10}}\\
\end{array}
\right)$\\
%%%%%%%%%%%%%%%%%%%%%%%%%%%%%%
\vspace{-3.5mm}
$A^{a}_{6,14}$\\
\end{tabular}
\end{center}
%%%%%%%%%%%%%%%%%%%%%%%%%%%%%%%%%%%%%%%%%%%%%%%%%%%%%%%%%%%%%%%%%%%%%%%%%%%%%%%%%%%%%%%%%%%%%%%%%%%%%%%%%%%%%%%%%%%%%%%%%%%%
\newpage
\vspace{10mm}
\begin{center}
\begin{tabular}{c c c}
\multicolumn{3}{c}{TABLE 3. (Continued)}\\
\hline \hline
Lie Algebra & Non Vanishing & Automorphism group \\&Structure Constants&\\
\hline \hline
%%%%%%%%%%%%%%%%%%%%%%%%%%%%%
\vspace{-3.5mm}
 $A^{a}_{6,18}$\\
\vspace{-3.5mm} $a\neq0$&$f^{3}_{12}=1,f^{4}_{13}=1,f^{6}_{14}=1$
& $\left(\begin{array}{cccccc}
a_{6} & a_{1} & a_{2} & a_{3} & a_{4} & a_{5} \\
-a a_{1} & a_{6} & a_{7} & a_{8} & a_{9} & a_{10} \\
0 & 0 & a_{6}^{2}+a a_{1}^{2} & a_{6}a_{7}+a a_{1}a_{2} & a_{1}a_{7}-a_{2}a_{6} & a a_{1}a_{3}+a a_{1}a_{9}+a_{6}a_{8}-a a_{4}a_{6}  \\
0 & 0 & 0 & a_{6}(a_{6}^{2}+a a_{1}^{2}) & a_{1}(a_{6}^{2}+a a_{1}^{2}) & a_{7}(a_{6}^{2}+a a_{1}^{2}) \\
0 & 0 & 0 & -a a_{1}(a_{6}^{2}+a a_{1}^{2}) & a_{6}(a_{6}^{2}+a a_{1}^{2}) & -a a_{2}(a_{6}^{2}+a a_{1}^{2}) \\
0 & 0 & 0 & 0 & 0 & (a_{6}^{2}+a a_{1}^{2})^{2}\\
\end{array}
\right)$ \\
& $,f^{5}_{23}=1,f^{6}_{25}=a$\\
%%%%%%%%%%%%%%%%%%%%%%%%%%%%%%%%%%%%%%%%%%
\vspace{-3.5mm}
$A_{6,19}$\\
\vspace{-3.5mm} &$f^{3}_{12}=1,f^{4}_{13}=1,f^{5}_{14}=1$ &
$\left(\begin{array}{cccccc}
a_{1} & a_{2} & a_{3} & a_{4} & a_{5} & a_{6} \\
0 & a_{1}^{3} & a_{7} & a_{8} & a_{9} & a_{10} \\
0 & 0 & a_{1}^{4} & a_{1}a_{7} & a_{1}a_{8} & a_{2}a_{7}+a_{1}a_{9}-a_{1}^{3}a_{3}  \\
0 & 0 & 0 & a_{1}^{5}& a_{1}^{2}a_{7} & a_{1}^{4}a_{2}+ a_{1}^{2}a_{8} \\
0 & 0 & 0 & 0 & a_{1}^{6} & a_{1}^{3}a_{7}  \\
0 & 0 & 0 & 0 & 0 & a_{1}^{7}\\
\end{array}
\right)$\\
\end{tabular}
\end{center}
%%%%%%%%%%%%%%%%%%%%%%%%%%%%%%%%%%%%%%%%%%%%%%%%%%%%%%%%%%%%%%%%%%%%%%%%%%%%%%%%%%%%%%%%%%%%%%%%%%%%%%%%%%%%%%%%%%%%%%%%%%%
\newpage
 \vspace{10mm}
\begin{center}
\begin{tabular}{c c c}
\multicolumn{3}{c}{TABLE 4: ad-invariant metric on all six dimensional solvable real Lie algebras }\\
  \hline
  \hline
Lie Algebra & Non Vanishing & ad-invariant metric \\&Structure Constants&\\
\hline \hline \vspace{-3.5mm}
$g_{6,23}$\\
\vspace{-4mm} $\alpha=0,\epsilon=0$
&$f^{1}_{23}=1,f^{3}_{26}=1,f^{4}_{36}=1$ &
$\left(\begin{array}{cccccc}
0 & 0 & 0 & 0 & 0 & g_{2} \\
0 & g_{1} & 0 & g_{2} & g_{3} & g_{4} \\
0 & 0 & -g_{2} & 0 & 0 & 0 \\
0 & g_{2} & 0 & 0 & 0 & 0 \\
0 & g_{3} & 0& 0 & g_{5} & g_{6}\\
g_{2} & g_{4} & 0 & 0 & g_{6} & g_{7}\\
\end{array}
\right)$\\
$,h=0 $\\
%%%%%%%%%%%%%%%%%%%%%%%%%%%%%%
\vspace{-3.5mm}
$g_{6,82}$\\
\vspace{-3.5mm} $\alpha=0$
&$f^{1}_{24}=1,f^{1}_{35}=1,f^{2}_{26}=\lambda,f^{3}_{36}=\lambda_{1}$
& $\left(\begin{array}{cccccc}
0 & 0 & 0 & 0 & 0 & -\lambda_{1}g_{1} \\
0 & 0 & 0 & \frac{\lambda_{1}g_{1}}{\lambda} & 0 & 0 \\
0 & 0 & 0 & 0 & g_{1} & 0\\
0 & \frac{\lambda_{1}g_{1}}{\lambda} & 0 & 0 & 0 & 0 \\
0 & 0 & g_{1} & 0 & 0 & 0 \\
-\lambda_{1}g_{1} & 0 & 0 & 0 & 0 & g_{2}\\
\end{array}
\right)$ \\
\end{tabular}
\end{center}
%%%%%%%%%%%%%%%%%%%%%%%%%%%%%%%%%%%
\newpage
\vspace{10mm}
\begin{center}
\begin{tabular}{c c c}
\multicolumn{3}{c}{TABLE 4. (Continued)}\\
  \hline
  \hline
Lie Algebra & Non Vanishing & ad-invariant metric \\&Structure Constants&\\
\hline \hline
%%%%%%%%%%%%%%%%%%%%%%%%%%%%%%%%%%%%
\vspace{-3.5mm}
$g_{6,90}$\\
\vspace{-3.5mm} $\alpha=0$
&$f^{1}_{24}=1,f^{1}_{35}=1,f^{4}_{26}=1,f^{5}_{36}=\nu_{0}$ &
$\left(\begin{array}{cccccc}
0 & 0 & 0 & 0 & 0 & -\nu_{0}g_{1} \\
0 & -\nu_{0}g_{1} & 0 & 0 & 0 & 0 \\
0 & 0 & g_{1} & 0 & 0 & 0 \\
0 & 0 & 0 & \nu_{0}g_{1} & 0 & 0 \\
0 & 0 & 0 & 0 & g_{1} & 0 \\
-\nu_{0}g_{1} & 0 & 0 & 0 & 0 & g_{2} \\
\end{array}
\right)$ \\
$\nu_{0}\neq1$ &$,f^{2}_{46}=1,f^{3}_{56}=-\nu_{0}$\\
%%%%%%%%%%%%%%%%%%%%%%%%%%%%%
\vspace{-3.5mm}
$g_{6,91}$\\
\vspace{-3.5mm}
&$f^{1}_{24}=1,f^{1}_{35}=1,f^{4}_{26}=1,f^{5}_{36}=1$ &
$\left(\begin{array}{cccccc}
0 & 0 & 0 & 0 & 0 & -g_{1} \\
0 & -g_{1} & 0 & 0 & 0 & 0 \\
0 & 0 & g_{1} & 0 & 0 & 0 \\
0 & 0 & 0 & g_{1} & 0 & 0 \\
0 & 0 & 0 & 0 & g_{1} & 0 \\
-g_{1} & 0 & 0 & 0 & 0 & g_{2} \\
\end{array}
\right)$ \\
\end{tabular}
\end{center}
%%%%%%%%%%%%%%%%%%%%%%%%%%%%%%%%%%%%%%%%%%%%%%%%%%%%%%%%%%%%%%%%%%%%%%%%%%%%%%%%%%%%%%%%%%%%%%%%%%%%%
\newpage
\appendix
\section{Appendix}
 \vspace{5mm}
 We give the automorphisms group for
$g_{6,92}(\alpha\neq0)$ , $g_{6,92}(\alpha=0)$ ,and $g_{6,93}(\alpha\neq0)$ Lie Algebras.\\
%%%%%%%%%%%%%%%%%%%%%%%%%%%%%%%%%%%%%%%
 \vspace{15mm}

$g_{6,92}(\alpha\neq 0)$\\
 \vspace{15mm}

\small{$\left(\begin{array}{cccccc}
\frac{-a_{2}a_{3}a_{5}^{2}\nu_{0}^{2}-a_{2}a_{3}a_{4}^{2}\mu_{0}\nu_{0}+a_{1}a_{5}^{3}\mu_{0}\nu_{0}+a_{1}a_{5}a_{4}^{2}\mu_{0}^{2}}{a_{5}^{2}\mu_{0}\nu_{0}}& 0 & 0 & 0 & 0 & 0 \\
\frac{2(2a_{2}a_{4}a_{7}\mu_{0}\nu_{0}+2a_{2}a_{5}a_{8}\mu_{0}\nu_{0}-2a_{1}a_{4}a_{9}\mu_{0}^{2}-2a_{1}a_{5}a_{10}\mu_{0}\nu_{0}+a_{2}a_{5}a_{7}\alpha\nu_{0}-a_{2}a_{4}a_{8}\alpha\mu_{0}-a_{1}a_{5}a_{9}\alpha\mu_{0}+a_{1}a_{4}a_{10}\alpha\mu_{0})}{a_{5}\mu_{0}(\alpha^{2}+4\mu_{0}\nu_{0})} & a_{1} & -\frac{a_{1}a_{4}}{a_{5}} & \frac{a_{2}\nu_{0}}{\mu_{0}} & -\frac{a_{2}a_{4}}{a_{5}} & 0 \\
\frac{2(a_{2}a_{4}a_{7}\alpha\nu_{0}+a_{2}a_{5}a_{8}\alpha\nu_{0}-a_{1}a_{4}a_{9}\alpha\mu_{0}-a_{1}a_{5}a_{10}\alpha\nu_{0}-2a_{2}a_{5}a_{7}\alpha\nu_{0}^{2}+2a_{2}a_{4}a_{8}\mu_{0}\nu_{0}+2a_{1}a_{5}a_{9}\mu_{0}\nu_{0}-2a_{1}a_{4}a_{10}\mu_{0}\nu_{0})}{a_{5}\nu_{0}(\alpha^{2}+4\mu_{0}\nu_{0})} & \frac{\mu_{0}a_{1}a_{4}}{\nu_{0}a_{5}}  & a_{1} & \frac{a_{2}a_{4}}{a_{5}}  & a_{2} & 0 \\
\frac{2(a_{5}^{2}a_{7}\alpha\nu_{0}-a_{4}a_{5}a_{8}\alpha\mu_{0}-a_{3}a_{5}a_{9}\alpha\nu_{0}+a_{3}a_{4}a_{10}\alpha\nu_{0}+2a_{4}a_{5}a_{7}\mu_{0}\nu_{0}+2a_{5}^{2}a_{8}\mu_{0}\nu_{0}-2a_{3}a_{4}a_{9}\mu_{0}\nu_{0}-2a_{3}a_{5}a_{10}\nu_{0}^{2})}{a_{5}\nu_{0}(\alpha^{2}+4\mu_{0}\nu_{0})}  & a_{3} & -\frac{a_{3}a_{4}}{a_{5}}&  a_{5}& -\frac{\mu_{0}a_{4}}{\nu_{0}} & 0 \\
\frac{2(-2a_{5}^{2}a_{7}\mu_{0}\nu_{0}+2a_{4}a_{5}a_{8}\mu_{0}^{2}+2a_{3}a_{5}a_{9}\mu_{0}\nu_{0}-2a_{3}a_{4}a_{10}\mu_{0}\nu_{0}+a_{4}a_{5}a_{7}\alpha\mu_{0}+a_{5}^{2}a_{8}\mu_{0}\alpha-a_{3}a_{4}a_{9}\mu_{0}\alpha-a_{3}a_{5}a_{10}\nu_{0}\alpha)}{a_{5}\mu_{0}(\alpha^{2}+4\mu_{0}\nu_{0})}  & \frac{a_{3}a_{4}}{a_{5}}  & \frac{a_{3}\nu_{0}}{\mu_{0}} & a_{4}& a_{5}& 0 \\
a_{6} & a_{7} & a_{8} & a_{9} & a_{10} & 1 \\
\end{array}
\right)$}\\

%%%%%%%%%%%%%%%%%%%%%%%%%%%%%%%%%%%%%%%%%%%%%
\vspace{15mm}

$\small{g_{6,92}(\alpha=0)}$\\
\vspace{15mm}

\small{$\left(\begin{array}{cccccc}
\frac{a_{2}(a_{3}^{2}\nu_{0}+a_{6}^{2}\mu_{0})}{a_{3}\nu_{0}}& 0 & 0 & 0 & 0 & 0 \\
a_{1} & a_{2} & -\frac{a_{2}a_{6}}{a_{3}} & 0 & 0 & 0 \\
\frac{a_{2}a_{9}a_{6}^{2}\mu_{0}+a_{1}a_{3}a_{6}\mu_{0}\nu_{0}+a_{2}a_{9}a_{3}^{2}\nu_{0}}{a_{3}^{2}\nu_{0}^{2}} & \frac{\mu_{0}a_{2}a_{6}}{\nu_{0}a_{3}}  & a_{2} & 0 & 0 & 0 \\
 A & \frac{a_{5}\mu_{0}}{\nu_{0}} & -\frac{\mu_{0}a_{5}a_{6}}{\nu_{0}a_{3}}&  a_{3}& -\frac{\mu_{0}a_{6}}{\nu_{0}} & 0 \\
a_{4} & \frac{a_{5}a_{6}\mu_{0}}{a_{3}\nu_{0}} & a_{5} & a_{6} & a_{3} & 0 \\
a_{7} & \frac{\mu_{0}(a_{2}a_{6}a_{8}a_{3}^{2}\nu_{0}+a_{2}a_{5}a_{9}a_{3}^{2}\nu_{0}+a_{2}a_{5}a_{9}a_{6}^{2}\mu_{0}+a_{1}a_{3}a_{5}a_{6}\mu_{0}\nu_{0}-a_{2}a_{4}a_{3}^{2}\nu_{0}^{2})}{a_{3}^{3}\nu_{0}^{2}a_{2}} & a_{8} & a_{9} & -\frac{\mu_{0}(a_{2}a_{6}a_{9}+\nu_{0}a_{1}a_{3})}{a_{2}a_{3}\nu_{0}} & 1 \\
\end{array}
\right)$}\\
\vspace{5mm}

\hspace{.5cm}A=\small{$\frac{a_{2}a_{8}a_{3}^{4}\nu_{0}^{2}+a_{2}a_{5}a_{6}a_{9}a_{3}^{2}\mu_{0}\nu_{0}+a_{1}a_{5}a_{3}^{3}\mu_{0}\nu_{0}^{2}+a_{2}a_{5}a_{9}a_{6}^{3}\mu_{0}^{2}+a_{1}a_{3}a_{5}a_{6}^{2}\mu_{0}^{2}\nu_{0}+a_{2}a_{8}a_{3}^{2}a_{6}^{2}\mu_{0}\nu_{0}-a_{2}a_{4}a_{6}a_{3}^{2}\mu_{0}\nu_{0}^{2}}{a_{3}^{3}a_{2}\nu_{0}^{3}}$}
%%%%%%%%%%%%%%%%%%%%%%%%%%%%%%%%%%%%%%%%%%%%
\newpage
\vspace{10mm}
$g_{6,93}(\alpha\neq0)$\\
\vspace{10mm}

$\left(\begin{array}{cccccc}
A & 0 & 0 & 0 & 0 & 0 \\
B & r & \nu_{0}(a_{3}-r) & a_{1} & \nu_{0}(a_{1}+a_{2}) & 0 \\
C & \nu_{0}(r-a_{3}) & a_{3} & \nu_{0}(a_{1}+a_{2}) & a_{2} & 0 \\
D & a_{1} & -\nu_{0}(a_{1}+a_{2}) & r & \nu_{0}(r-a_{3}) & 0 \\
E & -\nu_{0}(a_{1}+a_{2}) & a_{2} & \nu_{0}(a_{3}-r) & a_{3} & 0 \\
F & a_{5} & a_{6} & a_{7} & a_{8} & 1 \\
\end{array}
\right)$\\

\vspace{10mm}
\small{\hspace{-.5cm}A=$2a_{1}a_{2}\nu_{0}^{2}+2a_{2}^{2}\nu_{0}^{2}-a_{2}^{2}-2a_{3}^{2}\nu_{0}^{2}+2r\nu_{0}^{2}a_{3}+a_{3}^{2}$}\\

\vspace{5mm}
\small{\hspace{-.5cm}B=$2(\alpha^{3}a_{1}a_{5}+\alpha^{3}\nu_{0}a_{2}a_{6}-\alpha^{3}ra_{7}+2\alpha^{2}ra_{5}+8\nu_{0}^{4}a_{3}a_{5}+8\nu_{0}^{3}a_{3}a_{6}+8\nu_{0}^{4}a_{1}a_{7}+8\nu_{0}^{4}a_{2}a_{7}-8\nu_{0}^{3}a_{2}a_{8}-2\alpha^{2}a_{1}a_{7}
-8\nu_{0}^{4}ra_{5}+\alpha^{3}\nu_{0}a_{1}a_{6}-\alpha^{3}\nu_{0}a_{3}a_{8}+\alpha^{3}\nu_{0}ra_{8}-4\alpha\nu_{0}a_{2}a_{6}-4\alpha\nu_{0}^{2}a_{2}a_{5}+4\alpha\nu_{0}^{3}a_{1}a_{6}+4\alpha\nu_{0}^{3}a_{2}a_{6}-4\alpha\nu_{0}^{3}a_{3}a_{8}
+2\alpha^{2}\nu_{0}^{2}a_{3}a_{5}+2\alpha^{2}\nu_{0}^{2}a_{1}a_{7}+2\alpha^{2}\nu_{0}^{2}a_{2}a_{7}+2\alpha^{2}\nu_{0}a_{1}a_{8}-4\alpha\nu_{0}^{2}a_{3}a_{7}+4\alpha\nu_{0}a_{3}a_{8}+4\alpha\nu_{0}^{3}ra_{8}-2\alpha^{2}\nu_{0}^{2}ra_{5}+2\alpha^{2}\nu_{0}ra_{6})
/(8\alpha^{2}\nu_{0}^{2}+16\nu_{0}^{4}+\alpha^{4}-4\alpha^{2})$}\\

\vspace{5mm}
\small{\hspace{-.5cm}C=$2(\nu_{0}\alpha^{3}a_{1}a_{5}+\alpha^{3}\nu_{0}a_{2}a_{5}+\alpha^{3}\nu_{0}a_{3}a_{7}-\alpha^{3}\nu_{0}ra_{7}-4\alpha\nu_{0}a_{2}a_{5}-4\alpha\nu_{0}a_{3}a_{7}+4\alpha\nu_{0}^{2}a_{1}a_{6}
+8\nu_{0}^{2}\alpha a_{2}a_{6}-8\nu_{0}^{2}\alpha a_{3}a_{8}-2\alpha^{2}\nu_{0}^{2}a_{3}a_{6}+2\alpha^{2}\nu_{0}^{2}a_{2}a_{8}+4\nu_{0}^{3}\alpha a_{2}a_{5}+2\alpha^{2}\nu_{0}^{2}a_{1}a_{8}+4\alpha\nu_{0}^{3}a_{3}a_{7}+4\alpha\nu_{0}^{3}a_{1}a_{5}+2\alpha^{2}\nu_{0}^{2}ra_{6}-4\alpha\nu_{0}^{3}ra_{7}+4\alpha\nu_{0}^{2}ra_{8}+2\alpha^{2}\nu_{0}ra_{5}+\alpha^{3}a_{2}a_{6}-\alpha^{3}a_{3}a_{8}+4\alpha
a_{3}a_{8}+8\nu_{0}^{4}ra_{6}-8\nu_{0}^{4}a_{3}a_{6}+8\nu_{0}^{4}a_{2}a_{8}+8\nu_{0}^{4}a_{1}a_{8}-4\alpha a_{2}a_{6}+8\nu_{0}^{3}a_{3}a_{5}+8\nu_{0}^{2}a_{3}a_{6}+8\nu_{0}^{3}a_{2}a_{7}-8\nu_{0}^{2}a_{2}a_{8}-2\alpha^{2}\nu_{0}a_{1}a_{7})/(8\alpha^{2}\nu_{0}^{2}+16\nu_{0}^{4}+\alpha^{4}-4\alpha^{2})$}\\

\vspace{5mm}
\small{\hspace{-.5cm}D=$2(\nu_{0}\alpha^{3}a_{1}a_{8}+\alpha^{3}\nu_{0}a_{2}a_{8}-\alpha^{3}\nu_{0}a_{3}a_{6}+\alpha^{3}\nu_{0}ra_{6}+2\alpha^{2}\nu_{0}a_{1}a_{6}-2\alpha^{2}\nu_{0}^{2}a_{1}a_{5}-2\alpha^{2}\nu_{0}^{2}a_{2}a_{5}
-4\nu_{0}^{3}\alpha
a_{3}a_{6}-2\nu_{0}^{2}\alpha^{2}a_{3}a_{7}+4\alpha\nu_{0}^{2}a_{2}a_{7}+4\alpha\nu_{0}^{2}a_{3}a_{5}+4\nu_{0}^{3}\alpha
a_{1}a_{8}+4\nu_{0}^{3}\alpha
a_{2}a_{8}-4\alpha\nu_{0}a_{2}a_{8}+4\alpha\nu_{0}a_{3}a_{6}+2\alpha^{2}\nu_{0}ra_{8}+2\alpha^{2}\nu_{0}^{2}ra_{7}+4\alpha\nu_{0}^{3}ra_{6}+\alpha^{3}ra_{5}-\alpha^{3}a_{1}a_{7}-8\nu_{0}^{4}a_{3}a_{7}-8\nu_{0}^{4}a_{2}a_{5}-8\nu_{0}^{3}a_{2}a_{6}
+8\nu_{0}^{3}a_{3}a_{8}+2\alpha^{2}a_{1}a_{5}-8\nu_{0}^{4}a_{1}a_{5}-2\alpha^{2}ra_{7}+8\nu_{0}^{4}ra_{7})/(8\alpha^{2}\nu_{0}^{2}+16\nu_{0}^{4}+\alpha^{4}-4\alpha^{2})$}\\

\vspace{5mm}
\small{\hspace{-.5cm}E=$2(\nu_{0}\alpha^{3}a_{3}a_{5}+\alpha^{3}\nu_{0}a_{1}a_{7}+\alpha^{3}\nu_{0}a_{2}a_{7}-\alpha^{3}\nu_{0}ra_{5}+4\alpha\nu_{0}^{3}a_{3}a_{5}-4\alpha\nu_{0}^{2}a_{1}a_{8}-4\alpha\nu_{0}a_{3}a_{5}
+4\alpha\nu_{0}^{3}a_{1}a_{7}+4\alpha\nu_{0}^{3}a_{2}a_{7}-4\alpha\nu_{0}a_{2}a_{7}-2\alpha^{2}\nu_{0}^{2}a_{2}a_{6}-2\alpha^{2}\nu_{0}^{2}a_{1}a_{6}+2\nu_{0}^{2}\alpha^{2}a_{3}a_{8}+2\alpha^{2}\nu_{0}ra_{7}-2\alpha^{2}\nu_{0}a_{1}a_{5}
+8\alpha\nu_{0}^{2}a_{3}a_{6}-8\alpha\nu_{0}^{2}a_{2}a_{8}-4\alpha\nu_{0}^{3}ra_{5}-2\alpha^{2}\nu_{0}^{2}ra_{8}-4\alpha\nu_{0}^{2}ra_{6}+\alpha^{3}a_{3}a_{6}-\alpha^{3}a_{2}a_{8}-4\alpha
a_{3}a_{6}-8\nu_{0}^{4}a_{1}a_{6}-8\nu_{0}^{4}ra_{8}+8\nu_{0}^{2}a_{2}a_{6}-8\nu_{0}^{2}a_{3}a_{8}+4\alpha a_{2}a_{8}-8\nu_{0}^{4}a_{2}a_{6}+8\nu_{0}^{4}a_{3}a_{8}+8\nu_{0}^{3}a_{2}a_{5}+8\nu_{0}^{3}a_{3}a_{7})/(8\alpha^{2}\nu_{0}^{2}+16\nu_{0}^{4}+\alpha^{4}-4\alpha^{2})$}\\

\vspace{5mm}
\small{\hspace{-.5cm}F=$a_{4}$}\\

\vspace{5mm}
\hspace{-.5cm}$r=Rootof(z^{2}+a_{2}^{2}-a_{1}^{2}-a_{3}^{2})$\\
%%%%%%%%%%%%%%%%%%%%%%%%%%%%%%%%%%%
\endlandscape
%%%%%%%%%%%%%%%%%%%%%%%%%%%%%%%%%%%%%%%%%%%%%%%%%%%%%%%%%%%%%%%%%%%%%%%%%%%%%%%%%%%%%%%%%%%%%%%%%%%%%%%%%%%%%%%%%%%%%%%%%%%%
\newpage

{\bf Acknowledgments}

\vspace{3mm}

We would like to thank R.Campoamor-Stursberg, F. Darabi and GH.A.
Haghighatdust for carefully reading the manuscript and useful
comments.
%%%%%%%%%%%%%%%%%%%%%%%%%%%%%%%%%%%%%%%%%%%%%%%%%%%%%%%%%%%%%%%%%%%%%%%%%%%%%%%%%%%%%%%%%%%%%%%%%%%%%%%%%%%%%%%

\end{document}